\documentclass[twocolumn]{article}
\usepackage{mathtools}
\usepackage{graphicx}
\usepackage{color}
\usepackage{upgreek}
\usepackage{float}
\usepackage{txfonts}
\usepackage[T1]{fontenc}
\usepackage{longtable}
\usepackage{caption}
\usepackage[left=2cm,
            right=2cm,
            top=2cm,
            bottom=2cm]{geometry}
\usepackage[comma,super,sort&compress]{natbib}
\usepackage[colorlinks = true,
            linkcolor = blue,
            urlcolor  = blue,
            citecolor = blue,
            anchorcolor = blue]{hyperref}

\usepackage[noblocks]{authblk}

\usepackage{multirow}
\usepackage{siunitx}
\usepackage{xspace}
\newenvironment{Contfigure*}{%
\addtocounter{figure}{-1}%
\begin{figure*}}{%
\end{figure*}}

\newcommand{\rone}{FRB~20121102A\xspace}

\newcommand{\rsixseven}{FRB~20201124A\xspace}

\newcommand{\dmunit}{$\mathrm{pc~cm}^{-3}$\xspace}

\newcommand{\sfxc}{{\tt SFXC}\xspace}
\newcommand{\dspsr}{{\tt DSPSR}\xspace}
\newcommand{\psrchive}{{\tt PSRCHIVE}\xspace}
\newcommand{\digifil}{{\tt digifil}\xspace}
\newcommand{\filterbank}{{\tt filterbank}\xspace}
\newcommand{\sigproc}{{\tt SIGPROC}\xspace}
\newcommand{\presto}{{\tt PRESTO}\xspace}
\makeatother

\begin{document}

\title{Connecting repeating and non-repeating fast radio bursts via their energy distributions}

\author[1,2]{F.~Kirsten\thanks{Email: franz.kirsten@chalmers.se}}
\author[2,3]{O.~S.~Ould-Boukattine}
\author[4]{W.~Herrmann}
\author[5]{M.\,P.~Gawro\'nski}
\author[3,2]{J.\,W.\,T.~Hessels}
\author[6]{W.~Lu}
\author[2,3]{M.~P.~Snelders}
\author[3]{P.~Chawla}
\author[1]{J.~Yang}
\author[2]{R.~Blaauw}
\author[7]{K.~Nimmo}
\author[5]{W.~Puchalska}
\author[5]{P.~Wolak}
\author[3]{R.~van Ruiten}

\affil[1]{Department of Space, Earth and Environment, Chalmers University of Technology, Onsala Space Observatory, 439 92, Onsala, Sweden}
\affil[2]{ASTRON, Netherlands Institute for Radio Astronomy, Oude Hoogeveensedijk 4, 7991 PD Dwingeloo, The Netherlands}
\affil[3]{Anton Pannekoek Institute for Astronomy, University of Amsterdam, Science Park 904, 1098 XH, Amsterdam, The Netherlands}
\affil[4]{Astropeiler Stockert e.V., Astropeiler 1-4, 53902 Bad M\"{u}nstereifel, Germany}
\affil[5]{Institute of Astronomy, Faculty of Physics, Astronomy and Informatics, Nicolaus Copernicus University, Grudziadzka 5, 87-100 Toru\'n, Poland}
\affil[6]{Department of Astronomy and Theoretical Astrophysics Center, UC Berkeley, Berkeley, CA 94720, USA}        
\affil[7]{MIT Kavli Institute for Astrophysics and Space Research, Massachusetts Institute of Technology, 77 Massachusetts Ave, Cambridge, MA 02139, USA}

\maketitle

\section*{Summary} 

Fast radio bursts (FRBs) are extremely energetic, millisecond-duration radio flashes that reach Earth from extragalactic distances. Broadly speaking, FRBs can be classified as repeating or (apparently) non-repeating. It is still unclear, however, whether the two types share a common physical origin, differing only in their activity rate. Here we report on an unprecedented observing campaign that targeted one hyperactive repeating source, \rsixseven, for more than $2000~\mathrm{hr}$ using four $25-32\mathrm{-m}$ class radio telescopes. In total, we detect $46$ high-energy bursts, many more than one would expect given previous observations of lower-energy bursts using larger radio telescopes. We find a high-energy burst distribution that resembles that of the non-repeating FRB population, suggesting that apparently non-repeating FRB sources may simply be the rarest bursts from repeating sources. We also discuss how \rsixseven contributes strongly to the all-sky FRB rate and how similar sources would be observable even at very high redshift.

\section*{Introduction} 

The vast majority of FRB sources have only been detected once\cite{chime_2021_apjs}, but a small sub-population ($\sim\!2.6$\%) is known to burst repeatedly\cite{spitler_2016_natur, chimefrb_2023_repeaters}. Nonetheless, many apparent non-repeaters may be capable of repeating; in particular, recent results have shown that there is a wide range of FRB activity levels, with only very few sources being highly active \cite{chimefrb_2023_repeaters}. This suggests that the apparent non-repeaters are simply the least active FRB sources. However,  statistical studies have shown that the average emission bandwidth and burst duration differ between repeaters and non-repeaters \citep{chime_2021_apjs, pleunis_2021_apj}. This suggests that repeaters and non-repeaters have different origins, either a physically distinct progenitor or that a single type of source can produce different types of bursts. 

Probing the burst energy distribution of FRBs across many orders-of-magnitude, for both individual sources and the entire population, can help determine whether repeaters and non-repeaters have the same burst engines. 

Typically, FRBs have isotropic equivalent burst energies, $E$, in the range\cite{chime_2020_natur_galacticfrb} $E \sim 10^{36-41}~\mathrm{erg}$, though burst energies as high as $E = 2\times 10^{42}~\mathrm{erg}$ have been reported for distant sources\cite{ryder_2022_arXiv}. Telescope sensitivity strongly limits our ability to detect weaker bursts and on-sky time limits our ability to detect the rarest, most energetic bursts. In this paper, we will consider spectral energy density, $E_\nu = E/\nu$, where $\nu$ is the observed bandwidth of the emission. To convert quoted energies to spectral energies, we adopt a fiducial bandwidth of $300$~MHz as a `middle ground' between the statistically different bandwidths of repeating and apparently non-repeating FRBs\cite{pleunis_2021_apj}, unless specified otherwise. 

The cumulative burst spectral energy distribution of repeating FRBs can be modeled by a single or broken power-law\cite{li_2021_natur, zhang_2022_RAA}, $R(>E_\nu) \propto E_\nu^{\gamma}$, with the slope $\gamma$ sometimes steepening towards the high-energy tail, reaching values as extreme as $\gamma=-4.9$\cite{Kumar_2023_arXiv}. The measured value of $\gamma$ varies between sources and between observing epochs; e.g., for \rone, $\gamma$ ranges\cite{Jahns_2023_MNRAS, Hewitt_2022_MNRAS} between $-0.61$ and $-1.8$. There is also a hint that the burst energy distribution of \rone flattens\cite{Jahns_2023_MNRAS, Hewitt_2022_MNRAS} at the highest spectral energies ($E_\nu \gtrsim 3\times10^{31}~\mathrm{erg~Hz^{-1}}$), though the low number of observed events precludes a robust conclusion. 

In contrast, for apparently non-repeating FRBs, modeling of the population\cite{james_2021_mnras_frbstarformation,lu_2022_mnras,shin_2023_ApJ} shows a much flatter energy distribution, with $\gamma$ between about 0 to $-1$. Hyperactive repeating sources provide an opportunity to probe the high-energy burst distribution and to compare with apparently non-repeating sources. This is an important way to investigate whether repeaters and non-repeaters have the same progenitors. It also provides key input for FRB population simulations and applications for probing cosmology because the highest-achievable burst energies directly relate to the maximum distances from which we can observe FRBs.

One of the most hyperactive FRB sources to date is \rsixseven\cite{chime_2021_atel}, with $>3000$ bursts detected in $\sim25~\mathrm{hr}$ of observations with the FAST telescope\cite{wang_2022_atel}. In addition, the source has been reported to emit high-fluence bursts\cite{ouldboukattine_2022_atel_15190} that are detectable with relatively small $25-32\mathrm{-m}$ class radio telescopes. \rsixseven is in a region of enhanced star formation\cite{piro_2021_aa} in a massive star-forming galaxy at redshift\citep{fong_2021_apjl} $z = 0.098$. The known distance allows us to infer burst energies from measured fluences.

\section*{Observations \& Results}

\begin{figure*}
    \centering
    \includegraphics[width=\textwidth]{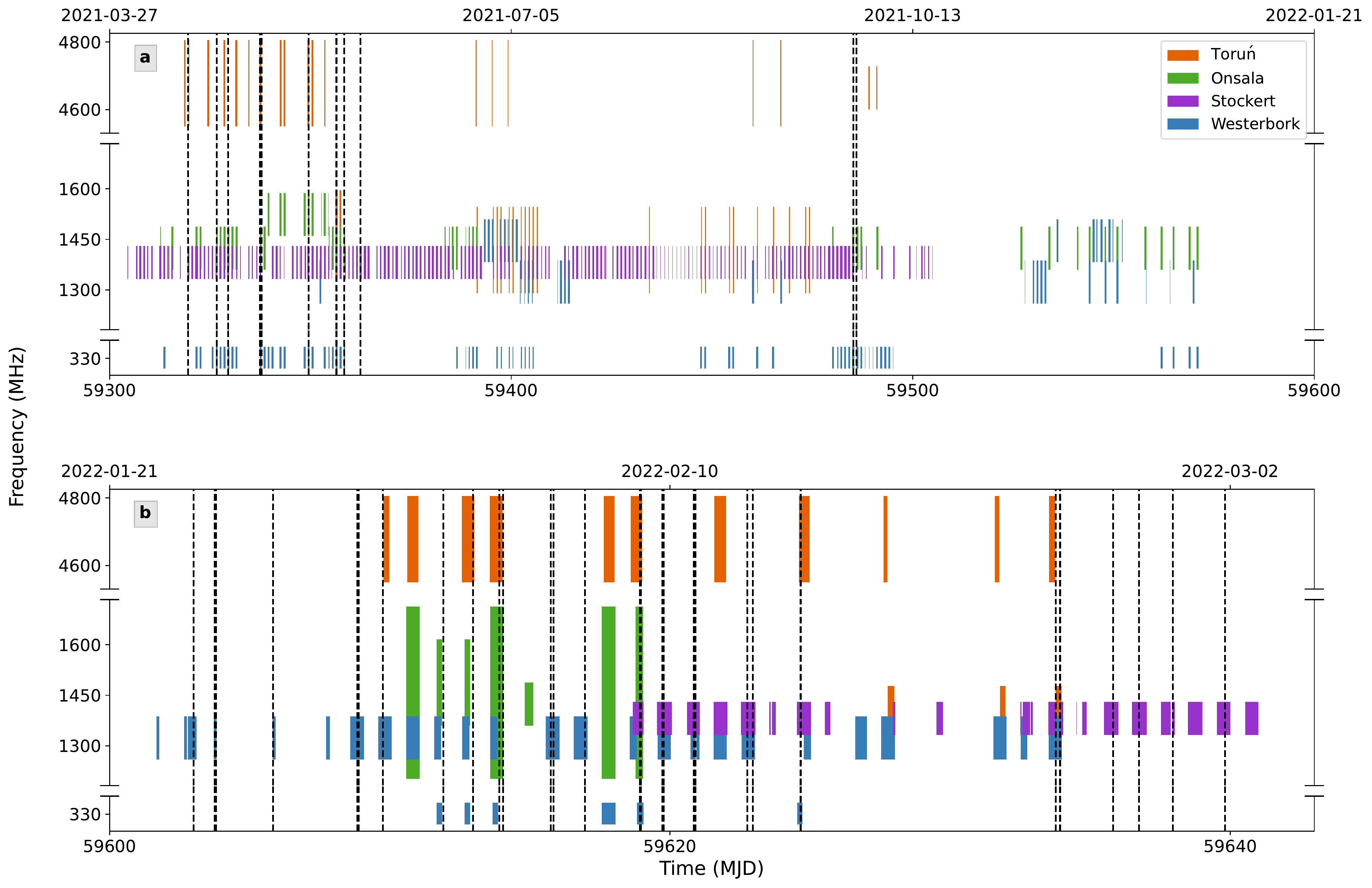}
    \caption{Overview of the \rsixseven monitoring campaign. Panel {\bf a} displays the MJD range $59300-59600$, while panel {\bf b} shows MJDs $59600-59645$.
    The corresponding calendar dates are listed on the top x-axes.
    Each coloured vertical block indicates an observation and associated telescope (annotated in the figure legend). The placement and extent along the y-axis indicates the observing band. Note that the frequency axis is discontinuous. The vertical black dashed lines denote detections of bursts from \rsixseven in the frequency range between $1202-1714~\mathrm{MHz}$ (despite spanning the entire frequency axis for visualisation purposes).}
    \label{fig:observations}
\end{figure*}

We observed \rsixseven between MJD~$59309$ and MJD~$59641$ (April~$2021$ -- March~$2022$; Figure~\ref{fig:observations}) for $2281~\mathrm{hr}$, spread over a one-year time span. Four radio telescopes were used: the $25-\mathrm{m}$ dish in Onsala, Sweden (O8); a $25-\mathrm{m}$ dish in Westerbork, The Netherlands (Wb); the $32-\mathrm{m}$ dish in Toru\'n, Poland (Tr); and the $25-\mathrm{m}$ telescope in Stockert, Germany (St). The observations were coordinated between the four telescopes with the aim to cover as broad a radio-frequency bandwidth as possible during contemporaneous observations while increasing the overall time on source (Table~\ref{tab:coverage}). During times of high source activity we observed for up to $12~\mathrm{hr}$ daily over the course of several weeks. We recorded raw voltages (amplitude and phase data) at Wb, O8, and Tr, while St recorded total-intensity data (Methods). The data were processed and searched for bursts using standard techniques and tools (Methods). 

\begin{figure*}
    \centering
    \includegraphics[width=\textwidth]{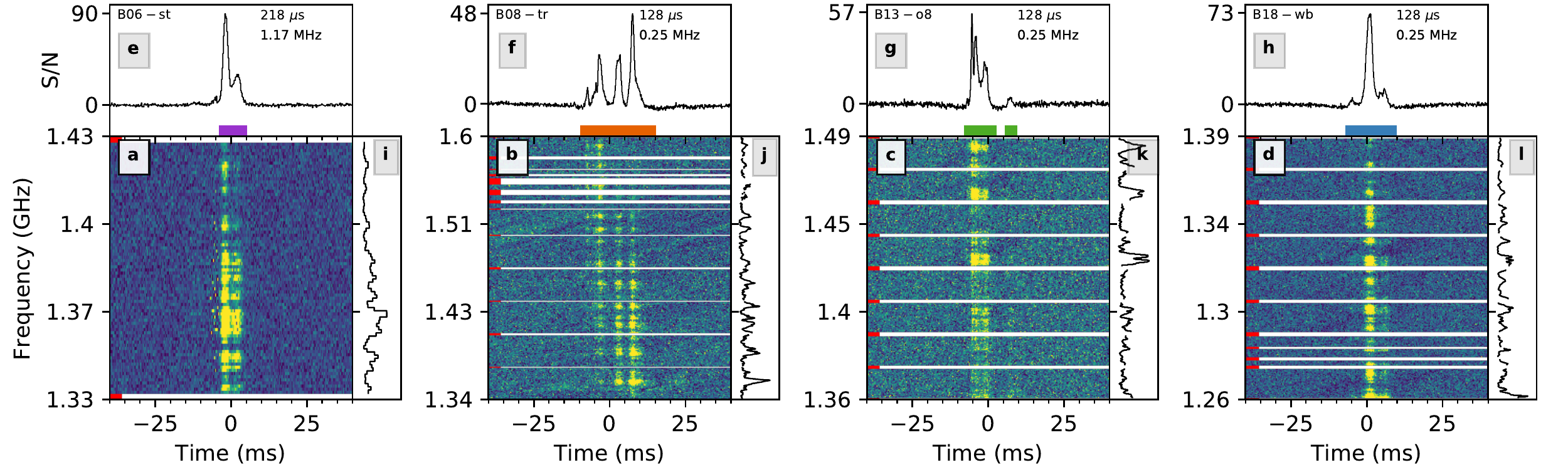}
    \caption{Subset of bursts detected in this campaign. Each panel consists of three sub-panels showing the dedispersed dynamic spectrum (panels {\bf a-d}; for Tr, O8, Wb coherently dedispersed using SFXC while incoherently dedispersed with {\tt DSPSR} for St as no baseband data were available), the frequency averaged time series (panels {\bf e-h}), and the spectrum of the time averaged `on-time' of the pulse (panels {i-l}). This on-time is indicated by the coloured horizontal bar under each time series, where the colour encodes the telescope at which the pulse was detected (from left to right): purple for St, orange for Tr, green for O8, and blue for Wb. The same on-time was used to compute the fluences in Table~\ref{tab:burst_properties}. In the time series panel, we indicate the pulse-ID in the top left corner, while the time and frequency resolution used in the plotting is indicated in the top right corner. In the dynamic spectra, horizontal white lines, with an additional red marker on the left, denote frequency ranges that were flagged because of RFI. For visualisation purposes the data were capped at the 98th percentile in the dynamic spectra. The full set of bursts is shown in Extended Data Figure~\ref{fig-extended:familyplot}.}
    \label{fig:subset-of-figs}
\end{figure*}

In total, we detected $46$ unique bursts in the frequency range $1200-1750~\mathrm{MHz}$ (L-band), after accounting for events that were detected by multiple telescopes simultaneously (Table~\ref{tab:numbers-per-epoch}). We consider sub-components to be part of a single burst if they are separated by no more than $100~\mathrm{ms}$ from each other. This choice is driven by the observed wait-time distribution\cite{xu_2022_natur, zhang_2022_RAA} of \rsixseven. Among our detections is the highest-fluence burst ever detected from this source (fluence $F \approx 1600~\mathrm{Jy~ms}$; Figure~\ref{fig:subset-of-figs}). Fourteen bursts were detected simultaneously at more than one dish. No  bursts were discovered at either P-band ($300-364~\mathrm{MHz}$) or C-band ($4550-4806~\mathrm{MHz}$) and none of the bursts discovered at L-band had a counterpart in the other frequency bands (Methods). The dynamic spectra, time series and spectra for a subset of bursts are shown in Figure~\ref{fig:subset-of-figs} (the full set is shown in Extended Data Figure~\ref{fig-extended:familyplot}).  Almost half of the bursts ($18/46$) are composed of $2$ or more components, most of which ($13/18$, i.e. $\sim70$\%) show the canonical `sad trombone' effect \cite{hessels_2019_apjl}, where burst emission at lower frequencies arrives later in time even after correcting for dispersive delay. We used the structure-optimising code {\tt DM-phase}\cite{seymour_2019_ascl} on one of our brightest multi-component bursts (B13-o8; Figure~\ref{fig:subset-of-figs}) to obtain a dispersion measure $\mathrm{DM}=410.8\pm0.3$~\dmunit (Methods); this DM is used throughout the rest of the analysis for all  bursts. The burst properties as listed in Extended Data Table~\ref{tab:burst_properties} were obtained from incoherently dedispersed data; except for the time-of-arrival (TOA), which was measured using coherently dedispersed data products where baseband data were available (i.e., for all bursts detected at Tr, Wb, O8; for St-bursts incoherently dedispersed data products were used for the TOAs; Methods). Particular care was given to the computation of the burst fluences because the large brightness and strong scintillation of the bursts systematically affects the recorded data through saturation effects (Methods; Extended Data Figures~\ref{fig-extended:2bit-demo} and \ref{fig-extended:fluence-ratio}). The median scintillation bandwidth, $\nu_s$, scaled to a canonical observing frequency of $1~\mathrm{GHz}$, $\nu_{s}^{1\mathrm{GHz}}=0.4\pm0.1~\mathrm{MHz}$, is consistent with previous observations\cite{main_2022_mnras}. Furthermore, we measure the characteristic temporal separation between individual burst components to be $\delta t=4.1^{+4.4}_{-2.1}~\mathrm{ms}$ (Extended Data Figure~\ref{fig-extended:wait-times}). This value is a factor $2-10$ larger than the one reported\cite{hessels_2019_apjl} for \rone.

\begin{figure*}
    \centering
    \includegraphics[width=\textwidth]{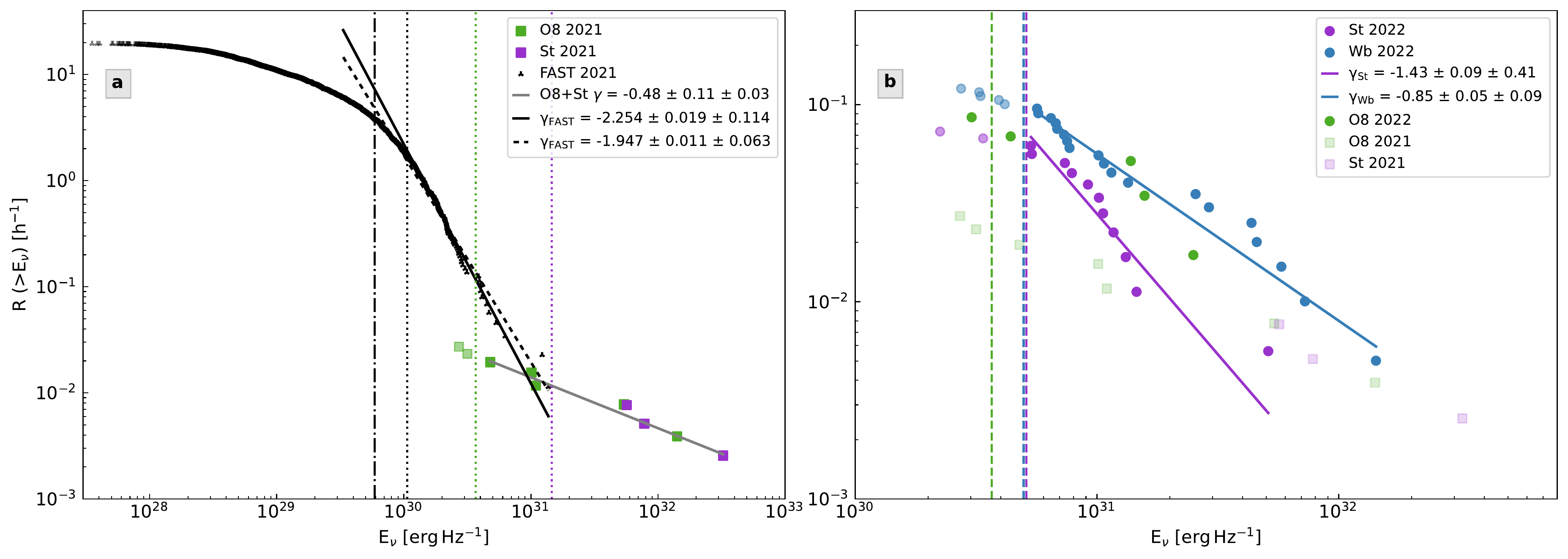}
    \caption{Burst spectral energy density distribution at different epochs. Panel {\bf a}: Cumulative burst rate distribution measured during the time range MJD $59305-59363$ in comparison to those reported by FAST during the same epoch \citep{xu_2022_natur}. The vertical green and purple dotted lines indicate the completeness threshold for O8 and St, respectively, beyond which we jointly fit the data with a simple power law (grey solid line). The vertical black dotted and dash-dotted lines indicate the thresholds beyond which the FAST data no longer follow a simple power law, as determined with the Python package \texttt{powerlaw}, and the break point of the broken power law as determined by \citet{xu_2022_natur}, respectively. We fit simple power laws to the data beyond the respective limits shown by the solid and dotted black lines. Pale coloured data points were excluded from the fit. Panel {\bf b}: Coloured circles indicate the burst rates observed after MJD~$59602$, as measured with Wb (blue), St (purple), and O8 (green). The coloured lines are the respective powerlaw fits. We did not fit the O8-data due to too few detections. Note that in this time range the St SEFD improved by a factor of $\sim3$ compared to the data taken in $2021$, lowering the completeness threshold compared to what is shown in the left panel. For comparison, we are also showing the burst rates as measured at O8 and St during the first activity window (pale coloured squares, same as left panel). The quoted uncertainties are composed of the statistical $1\sigma$-error of the least-squares fit (first error) and the $1\sigma$-uncertainty on the fitted slope from bootstrapping.}
    \label{fig:burstrates}
\end{figure*}

\begin{figure}
    \centering
    \includegraphics[width=1.0\columnwidth]{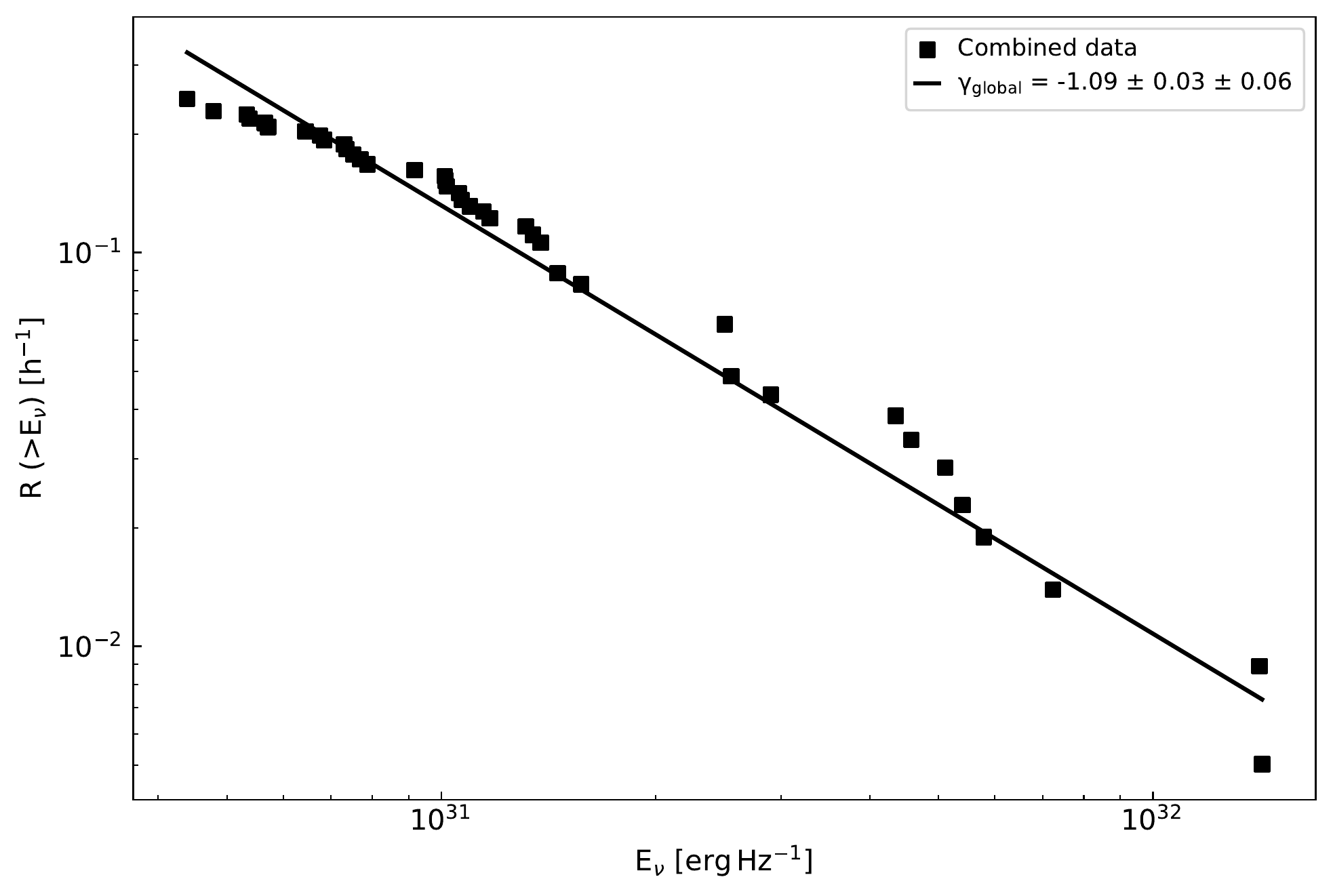}
    \caption{Burst energy distribution averaged over both activity windows and including almost all bursts detected at O8, Wb, and St. We exclude the data from St from the first activity window because of the large discrepancy in the detection threshold compared to the other stations. The solid black line denotes the best-fit power law.}
    \label{fig:discussion:global-rate}
\end{figure}

The vast majority of bursts was detected during two time ranges lasting roughly $60$ and $40$ days each (Figure~\ref{fig:observations} and Extended Data Table~\ref{tab:burst_properties}). In Figure~\ref{fig:burstrates} (left) we show the distributions of the burst spectral energy densities, $E_\nu$, during the first activity window compared with those observed with FAST during the same time range\cite{xu_2022_natur} (MJD~$59305-59363$). The energy distribution of our bursts shows no break, so we fit a simple power-law to the cumulative burst rate as function of spectral energy density, $R(>E_{\nu}) \propto E_{\nu}^{\gamma}$, using a least-squares technique (Methods). Jointly fitting the O8 and St data yields a power-law index $\gamma=-0.48\pm0.11\pm0.03$ (where the first error is the formal fitting uncertainty and the second error is derived from  bootstrapping; see Methods). This value is a factor $3$ lower (flatter) than that observed by FAST\cite{xu_2022_natur} for $E_\nu >5.9\times10^{29}~\mathrm{erg~Hz^{-1}}$ (using their median bandwidth of $185~\mathrm{MHz}$ to convert from energy to spectral energy), where $\gamma_2=-1.5\pm0.1\pm0.1$. However, the value of $\gamma_2$ was derived by fitting a broken power-law to the data split into energy bins\cite{xu_2022_natur} --- an approach that is not applicable for the relatively low number of bursts we detected. Therefore, for better comparability, we re-fit the FAST data following our technique (Methods) to find $\gamma_\mathrm{{FAST}} = -1.947\pm0.011\pm0.063$, which agrees with $\gamma_2$ within $2\sigma$. The bursts and the associated cumulative burst rates that we detected with O8, Wb, and St during the second activity window after MJD~$59602$ are shown in Figure~\ref{fig:burstrates} (right). We fit the data from Wb and St in the same way as described above and find power law slopes $\mathrm{\gamma_{St}=-1.43\pm0.09\pm0.41}$ and $\mathrm{\gamma_{Wb}=-0.85\pm0.05\pm0.09}$ for the St and Wb bursts, respectively. These energy distribution slopes are steeper than that from the first activity window, but are still much flatter for the Wb data than that of the high-energy tail in the FAST data\cite{xu_2022_natur}. The slope as found from the St-data during this epoch agrees with the one from Wb within the summed $1\sigma$ uncertainties but also agrees with $\gamma_2$ from Ref.\cite{xu_2022_natur}. Since the sensitivity thresholds of O8, Wb, and St during the second activity window are comparable, we can combine the data from all three telescopes from both activity windows (excluding the data from St during the first window as its detection threshold was a factor $\sim3$ higher at that time; see Methods) to get an estimate of the average slope $\mathrm{\gamma_{av}}$ of the energy distribution around an observing frequency of $1.3~\mathrm{GHz}$. Fitting the combined data yields $\mathrm{\gamma_{av}}=-1.09\pm0.03\pm0.06$ (Figure~\ref{fig:discussion:global-rate}). 

\section*{Discussion} 

\subsection*{\rsixseven's high-energy bursts are unlikely to be due to lensing effects}

We observe that the highest-energy bursts ($E_\nu \gtrsim 10^{31}~\mathrm{erg~Hz}^{-1}$) from \rsixseven occur at a much higher rate than expected based on the energy distribution of lower-energy bursts. One possible explanation is that burst brightness is boosted by propagation effects, such as lensing in an inhomogeneous plasma\cite{cordes_2017_apj} local to \rsixseven or an intervening gravitational potential\cite{Narayan_1993_LIACo}. However, if all bursts are affected by such a lens equally, the energy distribution would be shifted as a whole and no difference between the power-law slopes of low- and high-energy bursts would be seen. Moreover, in the limit of strong magnification with magnification factor $\mu >> 1$, the lensing cross section $\sigma(\mu) \propto \mu^{-2}$, independent of lensing potential\cite{oguri_2019_RPPh, Narayan_1993_LIACo}. Thus, if the brightest bursts that we see were caused by lensing, we would expect $\gamma \approx -2$, inconsistent with our measurements. In the case that amplification due to plasma lensing has a temporal dependence\cite{main_2018_natur}, one would expect a burst rate distribution that is variable in time. However, as shown in Figure~\ref{fig:observations}, we detect high-energy bursts throughout the entirety of the time range that was covered by FAST; there is no evidence for a variable energy distribution within the time range MJD~$59305-59363$. Thus, we conclude that the flattening of the burst energy distribution at high energies is most likely not a propagation effect. Rather, it more likely indicates a differing emission mechanism, emission site or beaming angle between low- and high-energy bursts. 

\subsection*{\rsixseven's high-energy burst distribution is similar to the population of non-repeaters} %

Burst rates as a function of energy have been reported for several repeating FRB sources, with the measured slopes of the fitted power-law distribution ranging from as steep as $-4.9$\cite{Kumar_2023_arXiv} to as flat as $-0.5$\cite{Hewitt_2022_MNRAS}. The slopes vary both as a function of time and as a function of the energy range that is considered. In general, the slope is apparently flatter towards lower energies\cite{Hewitt_2022_MNRAS, xu_2022_natur} ($E_\nu \lesssim 10^{29}~\mathrm{erg~Hz^{-1}}$). However, this often can be attributed to a lower completeness near the detection threshold of the telescope that is measuring the bursts. Nonetheless, in some cases, the low-energy turnover appears to be intrinsic to the emission process\cite{zhang_2022_RAA, xu_2022_natur, li_2021_natur}. The slope usually steepens towards the high-energy end ($E_\nu \gtrsim 3\times10^{29}~\mathrm{erg~{Hz^{-1}}}$) of the distribution \cite{zhang_2022_RAA} but a slight gradual flattening of the burst rates for very high-energy bursts ($E_\nu \gtrsim 10^{31}~\mathrm{erg~Hz^{-1}}$) has been observed for at least one source\cite{Hewitt_2022_MNRAS, Jahns_2023_MNRAS}. Here we observe a strong flattening of the burst rate slope in a previously unexplored ultra-high-energy range for \rsixseven.

Unaffected by the limited bandwidth of our observations (Methods), the average high-energy slope that we measure (Figure~\ref{fig:discussion:global-rate}) broadly agrees with that found for non-repeating FRBs\cite{shin_2023_ApJ}, and our value also agrees with various empirical models\cite{james_2021_mnras_frbstarformation, lu_2022_mnras} that find a power law slope $\gamma\approx-1$ for non-repeating FRBs. This indicates that the distribution of the ultra-high-energy tail ($\mathrm{E_\nu \gtrsim 10^{31}erg~Hz^{-1}}$) of \rsixseven bursts resembles that of non-repeaters, while the low-energy bulk of the bursts does not. Thus, we are seeing both repeater- and non-repeater-like behaviour from a single FRB source. 

In the case of apparently non-repeating FRBs, we may be observing bursts from a similar high-energy tail as what we report here for \rsixseven. Since such bursts are extremely rare, occurring once per hundreds to thousands of hours, or more (we see no high-energy cut-off in the burst energy distribution), such a source would appear to be non-repeating. 

However, \rsixseven's properties in terms of burst morphology, width and bandwidth are currently statistically indistinguishable between the low- and high-energy end of the distribution. This contradicts what has been shown for the repeater and non-repeater populations as a whole\cite{pleunis_2021_apj}. A careful comparison of, e.g., the polarimetric properties of high-energy bursts from \rsixseven with those of non-repeaters can shed further light on their possible connections. If non-repeaters are indeed drawn from the high-energy tail of a repeater's burst energy distribution, approximately $1000~\mathrm{hr}$ of observing time, or more, is required in order to detect a repeat burst from a given source. Very few FRB sources have been observed in this way. On the other hand, in case the observed flattening is a peculiarity of \rsixseven --- and if the overall population of FRBs generally follows the steeper rate distribution as observed by FAST (and assuming all FRBs repeat) --- we can extrapolate this rate to the spectral energy density of our brightest burst, B06-st. Assuming a similar energy distribution as that observed by FAST, the rate of our brightest burst B06-st is approximately $\mathrm{10^{-6}~hr^{-1}}$.

\subsection*{\rsixseven contributes strongly to the overall sky rate}

The ASKAP FRB-survey\cite{shannon_2018_natur} was conducted in a similar frequency band as our observations. Based on modelled number counts\cite{lu_2019_ApJ} for $20$ ASKAP bursts, $N(>F)=20(F/50~\mathrm{Jy~ms})^{-1.5}$, and the total exposure ($5.1\times10^5~\mathrm{deg^{2}~hr}$) of the ASKAP survey, an FRB all-sky rate above a fluence $F>100\mathrm{~Jy~ms}$, $\mathrm{R_{sky}}(F>100~\mathrm{Jy~ms})=5\times10^3~\mathrm{sky^{-1}~yr^{-1}}$, was reported\cite{shannon_2018_natur}. During our campaign, we found $17$ bursts from \rsixseven above this threshold which, considering the amount of time spent on source, corresponds to $1.3$\% of $\mathrm{R_{sky}}(F>100\mathrm{~Jy~ms})$. Moreover, considering only the $3$ brightest bursts in our sample, i.e the ones with $F>500\mathrm{~Jy~ms}$, the contribution of \rsixseven to $\mathrm{R_{sky}}(F>500\mathrm{~Jy~ms})$ during its active state is even twice as high, at $2.6$\%. Since the spectral energy distribution that we observe is shallower than the assumed power-law slope of $-1.5$ used in the modelled number counts\cite{lu_2019_ApJ}, these fractional contributions are only a lower limit. This demonstrates that hyperactive repeaters can account for a significant fraction of all observed FRBs.

\subsection*{\rsixseven would be observable even at very high redshift}

The emission from FRB sources is subject to dispersion, scattering, and Faraday rotation while travelling through the cold plasma of instellar and intergalactic space. As such, the signals carry the imprints of matter density structure along a particular line-of-sight, allowing measurements of, e.g., galaxy halo matter densities\cite{prochaska_2019_sci}. By providing a complete measurement of the free electrons along the path, via the DM, FRBs have already been used to trace the diffuse intergalactic medium and, thereby, contribute to solving\cite{macquart_2020_natur} the `missing baryon problem'. Furthermore, it was  suggested\cite{wucknitz_2021_aa} to use individual well-localised, gravitationally lensed repeating FRBs to measure the Hubble constant via changing delays between the arrival times of lensed versions of subsequent bursts. As such, FRBs are proven cosmological probes with great future potential as well. The question of how far back in cosmic history we can probe the Universe with FRBs is directly related to their energetics.

Assuming a burst bandwidth of $\mathrm{1~GHz}$, recent empirical models of the FRB population\cite{shin_2023_ApJ} find a characteristic spectral energy cut-off of $\mathrm{E_\nu^{char}=2.38^{+5.35}_{-1.64}\times10^{32}~erg~Hz^{-1}}$. The most energetic burst that we report here, B06-st, had an isotropic equivalent spectral energy of $\mathrm{E_\nu = 3.1^{+0.6}_{-0.7}\times10^{32}~erg~Hz^{-1}}$, only slightly below the upper limit of the estimated $\mathrm{E_\nu^{char}}$. For the FAST data shown in Figure~\ref{fig:burstrates}, the completeness threshold\cite{xu_2022_natur} was $53~\mathrm{mJy~ms}$. Thus, given the spectral energy that we measure for B06-st, such a burst would have been observable out to redshift $z=12.9^{+1.4}_{-1.5}$ with a telescope such as FAST. 

At that redshift, the emitted frequency of the pulse would need to be $\sim 19~\mathrm{GHz}$ to be observable at $1.4~\mathrm{GHz}$ from Earth. This is plausible, considering that \rone has been detected at $8~\mathrm{GHz}$ \citep{gajjar_2018_apj}, which corresponds to an emission frequency of $\sim 10~\mathrm{GHz}$ at the source given its redshift\cite{tendulkar_2017_apjl} $z=0.19273$. Similarly, in order to be observable in the CHIME band ($400-800~\mathrm{MHz}$), pulses emitted at redshift $z=12.9$ would need to be emitted at $\sim 8~\mathrm{GHz}$ at the source. Employing the Macquart-relation\cite{macquart_2020_natur}, such a high redshift would imply a DM $\sim 5700$~\dmunit. This is close to double the DM seen from any other FRB\cite{chime_2021_apjs}, to date, and the largest measured FRB redshift\cite{ryder_2022_arXiv} is currently $z=1.016\pm0.002$. Nevertheless, purely from an energetics point-of-view, it is not implausible to expect hyperactive FRBs like \rsixseven at redshifts beyond $3$, allowing for novel cosmological studies using repeat bursts. However, it has been pointed out \cite{ocker_2022_apj} that at such high redshifts individual bursts might be subject to scatter broadening as large as $300~\mathrm{ms}$ at $1~\mathrm{GHz}$ due to intervening galaxies. This would reduce the detectability of high-redshift FRBs.

\section*{Conclusions}

Using four small $25-32\mathrm{-m}$ class radio telescopes, we present a multi-frequency observing campaign targeting the hyperactive, repeating \rsixseven. This campaign is unprecedented in terms of the amount of observing time spent on source, and we conclude that ultra-high-energy ($E_\nu \gtrsim 10^{31}~\mathrm{erg~Hz}^{-1}$) bursts occur much more frequently than would have been expected based on previous observations of lower-energy bursts\cite{xu_2022_natur}. We argue that our detected bursts are intrinsically higher in energy, as opposed to being boosted in brightness by propagation effects like lensing. 

Given our results, \rsixseven generates bursts spanning at least $6$ orders-of-magnitude in spectral energy density, a similar span to the Galactic magnetar SGR~1935+2154\cite{kirsten_2021_natas} but reaching to much higher energies. Moreover, the burst energy distribution flattens towards the highest-observed energies. This high-energy distribution resembles that seen from the population of non-repeating FRBs as a whole, suggesting that apparently non-repeating FRBs may simply be sampling the rare high-energy events of sources that are capable of repeating. The highest-energy bursts may originate from a separate emission mechanism or emission region at the progenitor source. Further evidence for such a scenario might be found by comparing the polarimetric properties of low- and high-energy bursts.

We also showed that \rsixseven's extremely energetic bursts, though rare, mean that a similar hyperactive repeating source would be detectable out to very high ($z=12$) redshift using the world's most sensitive radio telescopes, though large on-sky time would be necessary to detect repeat bursts.

Lastly, the large number of high-fluence ($F>500~\mathrm{Jy~ms}$) bursts that we detected from \rsixseven constitute a significant fraction (at least $\sim2.6$\%) of the estimated all-sky FRB rate\cite{lu_2019_ApJ, shannon_2018_natur}. This shows that, in this high-fluence range, the all-sky rate is significantly influenced by a small number of hyperactive sources. 

\clearpage
\newpage
\section*{Methods} 
\subsection*{Observations and data reduction}
\subsubsection*{Onsala (O8), Westerbork (Wb), and Toru\'n (Tr)}
At O8, Wb, and Tr we recorded raw voltages (`baseband data', providing both the amplitude and phase of the electromagnetic signal) in VDIF format\cite{whitney_2010_ivs} using the local DBBC2 and Flexbuff systems. Both left and right circular polarisations were sampled at the Nyquist rate and recorded as $2$-bit samples. Depending on observing frequency and setup, the recorded bandwidth varied between $64~\mathrm{MHz}$ and $512~\mathrm{MHz}$, divided into subbands of $8$, $16$, or $32~\mathrm{MHz}$ (Table~\ref{tab:coverage}). At O8 all observations were conducted at L-band (between $1202-1714~\mathrm{MHz}$, with varying bandwidths used). Wb covered P-band ($300-364~\mathrm{MHz}$) when co-observing with O8 and L-band when observing on its own. Tr observed at C-band ($4500-4800~\mathrm{MHz}$, with varying bandwidths used) when co-observing with another dish, or otherwise also covered a part of L-band. For the exact frequency ranges and observed hours per station, see Table~\ref{tab:coverage}.

All of the data from Wb and O8 and a subset of the data from Tr were transferred via the internet to a dedicated processing machine, `ebur', at Onsala Space Observatory in Sweden. The processing pipeline\cite{kirsten_2021_natas} performs the following steps:
\begin{itemize}
    \item generate total intensity (Stokes I) filterbanks ($8$-bit encoding) of varying time and frequency resolution from the baseband data,
    \item search the filterbank data for bursts using \href{https://sourceforge.net/projects/heimdall-astro/}{{\tt Heimdall}},
    \item classify the candidates as found by Heimdall using the machine learning classifier FETCH\cite{agarwal_2020_mnras},
    \item manually inspect the diagnostic plots generated by FETCH.
\end{itemize}
The time and frequency resolution of the total intensity filterbanks varied depending on observing frequency: $1.024~\mathrm{ms}$ / $7.8125~\mathrm{kHz}$ at P-band, $0.128~\mathrm{ms}$ / $62.5~\mathrm{kHz}$ at L-band, and $0.064~\mathrm{ms}$ / $500.0~\mathrm{kHz}$ at C-band. These values were chosen as a compromise between the Nyquist limit and the maximal residual DM-smearing in the lowest frequency channel of each observing band. In the Heimdall search, the DM range over which the search was conducted was limited to $\mathrm{DM_{FRB}\pm 50}$~\dmunit, where $\mathrm{DM_{FRB}\cite{chime_2021_atel}=413.0}$~\dmunit. For radio frequency interference (RFI) mitigation we implemented a static frequency mask that was obtained via manual inspection of parts of the data. The detection threshold was set to $7\sigma$, which translates to fluence limits of roughly $42~\mathrm{Jy~ms}$, $7~\mathrm{Jy~ms}$, and $3~\mathrm{Jy~ms}$ at P-, L- and C-band, respectively. When classifying the Heimdall-generated burst candidates as either astronomical signals or RFI, we use FETCH-models\cite{agarwal_2020_mnras} a and h with a probability threshold of $0.5$. 

\subsubsection*{Stockert (St)}\label{sec:observations:stockert}

The recording setup and burst detection pipeline at St were developed independently from the one for O8/Wb/Tr and hence use different tools and detection thresholds. At St, we observed in the frequency range $1332.5-1430.5~\mathrm{MHz}$, recording total intensity $32$-bit data with a Pulsar Fast Fourier Transform\cite{barr_2013_mnras} (PFFTS) backend. The data were initially stored as `PFFTS' files, which is the instrument's specific format. These were subsequently converted to the standardised filterbank format using the tool \filterbank which is part of the \sigproc\cite{lorimer_2011_ascl} package. The resulting data had time and frequency resolution of $218.45~\mu$$\mathrm{s}$ and $586~\mathrm{kHz}$ and were stored as $32$-bit floats. We used the tools \texttt{rfifind}, \texttt{prepsubband} and \texttt{single\_pulse\_search} from the \presto\cite{ransom_2011_ascl} package to subsequently remove RFI, incoherently dedisperse the data (using $\mathrm{DM=412.0}$~\dmunit) and search for bursts. Between $2021$ November $24$ and $2022$ February $8$, the St dish underwent a significant upgrade, reducing the system equivalent flux density (SEFD) by almost a factor of three (Table~\ref{tab:coverage}), thereby improving the sensitivity of the dish significantly. In the pulse searches we used a detection threshold of $8\sigma$, which corresponds to fluence limits of $21$ and $7~\mathrm{Jy~ms}$ before and after the upgrade, respectively.

\subsubsection*{Digitisation artefacts}\label{sec:digitisation_artefacts}

As discussed above, we recorded $2$-bit baseband data at O8, Wb, and Tr. The limited dynamic range of these samples in combination with the large fluences of some of the bursts ($\mathrm{S/N\approx100}$) leads to digitisation artefacts in the data whereby power is `scattered' in time and frequency\cite{jenet_1998_pasp}. Most notably this manifests as `depressions' in the data around the times of the bursts (e.g., bursts B08-o8, Extended Data Figure~\ref{fig-extended:2bit-demo}). A similar digitisation artefact was also found and described\cite{Ikebe_2023_PASJ} for a burst from \rsixseven detected at $2.2~\mathrm{GHz}$. As a result, the measured overall fluences will be underestimated. In order to quantify this effect and compensate for it, we measure the fluences from data products generated from the baseband data in three different ways. Case~i: we create coherently dedispersed total intensity filterbanks with the Super FX Correlator \sfxc\cite{keimpema_2015_exa}. These filterbank data are subsequently converted to `archive' files with the standard pulsar software package \dspsr\cite{vanstraten_2011_pasa}. Case~ii: we use the tool \digifil that is part of \dspsr to generate total intensity filterbanks (no dedispersion applied) which we then convert to archives with \dspsr (applying incoherent dedispersion); and Case~iii: we apply a `scattered power correction' (SPC) algorithm\cite{vanstraten_2013_apjS} that is part of \psrchive\cite{vanstraten_2012_art, hotan_2004_pasa} to the archives generated in Case~ii. 

The reason for implementing Case~iii is that \digifil applies a $2$-bit correction to the baseband data\cite{jenet_1998_pasp} that effectively over-compensates for the limited dynamic range in the case of bright bursts. The effect is that the overall measured fluence is over-estimated compared to the real fluence. Extended Data Figure~\ref{fig-extended:2bit-demo} shows a comparison of the burst profiles for B08-o8 generated in these three ways. The profile generated via \sfxc dips below the mean of the off-pulse region in between components while the regular \digifil generated profile stays above the mean. Even though this could be intrinsic to the pulse, the fact that the profile appears to begin/rise early compared to the \sfxc profile indicates that there is an overall `skirt' around the burst that is due to the $2$-bit correction implemented in \digifil. The profile generated from the archive corrected with the SPC-algorithm (i.e. Case~iii) does not show this skirt while also not dropping below the mean in between components. Thus we conclude that this data product provides the best estimate of the true fluence of the burst. Extended Data Figure~\ref{fig-extended:fluence-ratio} shows a relative comparison of the measured fluences scaled by the SPC-corrected fluences. Overall, compared to the Case~iii fluences, the Case~ii fluences (uncorrected \digifil) are $5-10$\% higher, while the Case~i fluences (\sfxc-generated) are lower by a factor $1.5-2.0$. There is no clear fluence threshold above which the quantisation limitation shows its effects. Rather, it is the range in time-frequency space over which power is concentrated that determines whether or not `depressions' occur. A single bright scintil can result in scattered power while the overall band-integrated fluence is not higher compared to other bursts. While making these corrections is important for achieving the most accurate fluences possible, we note that factor of $2$ uncertainties in the burst energies do not significantly affect the conclusions that we draw in this paper.

For the data recorded at St, the $2$-bit quantisation limitations do not apply and we measure the burst properties directly from the recorded filterbanks and the archive files generated via Case~ii.

\subsection*{Analysis \& Results}

\subsubsection*{DM-optimisation}
We used the structure-optimizing code {\tt DM-phase} \citep{seymour_2019_ascl} on one of our brightest multi-component bursts (B13-o8, Figure~\ref{fig:subset-of-figs}) to obtain $\mathrm{DM}=410.8\pm0.3$~\dmunit ({\tt DM-phase} uses a DM constant of $1/0.000241$ $\mathrm{GHz^2~cm^3~pc^{-1}~\mu s}$). To assess any potential evolution of the DM over the roughly one-year time span of our observing campaign, we ran {\tt DM-phase} on $9$ other bright, multi-component bursts detected throughout the campaign. We did not measure any significant evolution, nor any DM values that disagree beyond their respective $2\sigma$ uncertainties with our initial measurement (Extended Data Table~\ref{tab:bright-ten-bursts}). Any DM offsets on the order of our measurement uncertainties have only negligible impact on the values listed in Extended Data Table~\ref{tab:burst_properties} (the most notable would be a difference in the time-of-arrival (TOA) of $\sim1.2~\mathrm{ms}$ for a DM offset of $0.5$~\dmunit). Thus, given the challenges in measuring the `correct' DM of an FRB\cite{Nimmo_2023_MNRAS}, we used this DM value in the rest of the analysis on all detected bursts.

\subsubsection*{The effect of bandwidth on the number of detected bursts}
During our campaign, we detected bursts spanning more than an order of magnitude in spectral energy $E_\nu$. A power law slope similar to the one observed with FAST\cite{xu_2022_natur} would require a difference in rate of a factor of $\gtrsim10^2$ between the high- and low-energy bursts, i.e. a factor $\gtrsim10$ larger than what we find. Previous studies\cite{Hewitt_2022_MNRAS} on \rone report that their low-energy bursts ($E_\nu\lesssim3\times10^{29}~\mathrm{erg~Hz^{-1}}$) are typically more narrow in bandwidth than their high-energy bursts. In our case this could imply that due to our limited bandwidth of $\sim100~\mathrm{MHz}$, we might miss low-energy bursts as they might be bright outside of our observing band while the high-energy bursts are sufficiently broad-band to always fall within our frequency window. Conversely, the FAST L-band receiver covers a frequency range of $1.0-1.5~\mathrm{GHz}$ and the median bandwidth\cite{xu_2022_natur} of the bursts ($185~\mathrm{MHz}$) indicates that bursts with different central frequencies appear in different parts of the FAST L-band where they might be outside our limited frequency range. However, the cumulative burst rate that we derive for our `low-energy' bursts ($E_\nu\sim10^{31}\;\mathrm{erg~Hz^{-1}}$) is in good agreement with similarly energetic bursts detected with FAST\cite{xu_2022_natur} in the same time range. Therefore, we exclude the possibility of missing a large fraction of bursts because of a frequency `windowing` effect. Even if we miss some of the low-energy bursts because of RFI, we deem it highly unlikely that we miss $90$\% of the possible low-energy bursts. Instead, we conclude that we detect a factor of $\sim10$ `too many' bursts compared to what one would expect from the FAST-slope\cite{xu_2022_natur} in the spectral energy range $\sim10^{30}-10^{32}\mathrm{~erg~Hz^{-1}}$ during this time period. 

\subsubsection*{Burst properties}

Except for the TOA, all burst properties listed in Extended Data Table~\ref{tab:burst_properties} were obtained from the archive files that were corrected with the SPC algorithm (Case~iii above). First, we manually determine a flagging mask using the \psrchive tools \texttt{pazi} and \texttt{psrzap}. This mask excises RFI and discards $5$\% of bandwidth at the top and bottom of each subband. We subsequently downsample the data to a time and frequency resolution of $256~\mu$$\mathrm{s}$ and $500~\mathrm{kHz}$, respectively, and inspect the dedispersed burst profiles by eye and manually select the start and stop bins of burst components. For each of these (component-based) time ranges, we compute the $2$D-autocorrelation function in time and frequency. The resulting autocorrelation spectra and time series are fitted separately with a $1$D-Gaussian to determine the component width in time (the `time-Gaussian') and in frequency (the `frequency-Gaussian'). In order to measure a scintillation bandwidth for each burst component, we first subtract the frequency-Gaussian from the autocorrelation spectra and then fit a Lorentzian to the central component of the residuals. The half-width at half-maximum of this Lorentzian is an estimate of the scintillation bandwidth $\nu_s$\cite{rickett_1990_araa}. We list these values scaled to a frequency of $1~\mathrm{GHz}$ in Extended Data Table~\ref{tab:burst_properties}. For the scaling of $\nu_s$ with central frequency $\nu_c$, $\nu_s\propto\nu_{c}^\alpha$, we adopted a canonical power-law slope $\alpha=4.0$. The median of $\nu_{s}^{1\mathrm{GHz}}=0.4\pm0.1~\mathrm{MHz}$ is consistent with the earlier measurements\cite{main_2022_mnras}. Finally, we take the burst frequency extent to be the total observing bandwidth unless twice the full width at half maximum (FWHM) of the frequency-Gaussian is less than $75$\% of the total bandwidth (BW, i.e. $\mathrm{2 \times FWHM < 0.75 \times BW}$), in which case we by eye determine the burst frequency extent. This approach is chosen to ensure no flux is lost due to flagged channels (which in either case would lead to an underestimated FWHM of the frequency-Gaussian). The fluences are computed by first determining the flux densities per time bin via the radiometer equation\cite{cordes_2003_apj} and subsequently summing over the manually selected on-time region. From the fluences we then determine the isotropic-equivalent spectral luminosity of the bursts using the known luminosity distance\cite{ravi_2022_mnras}  $D_L=453\pm0.1~\mathrm{Mpc}$ ($z = 0.098$) of \rsixseven.

\subsubsection*{Burst times-of-arrival}
During the recording it is not uncommon that some data loss occurs (on the order of a few seconds accumulated over a $15$-minute recording). \digifil has currently no functionality to detect and account for such data loss, which eventually leads to inaccurate time stamps in the generated filterbanks.  Having been developed for correlating interferometric observations, SFXC has the required functionality to compensate for such losses. Therefore we make use of SFXC to generate coherently dedispersed filterbanks (SFXC uses the same DM constant as {\tt DM-phase}) to compute the TOA of each burst component. We fit multi-component Gaussians to the frequency-averaged profile (time resolution of $64~\mu$$\mathrm{s}$) of each burst and take the centre of each Gaussian component as the component's TOA. The time stamps in the SFXC-generated filterbanks are referenced to the geocenter and the reference frequency used for coherently dedispersing the pulses is the centre frequency of the highest subband. In Extended Data Table~\ref{tab:burst_properties}, the listed TOAs are referenced to the barycenter (in the TDB timescale) at infinite frequency. The times designate either the middle between the first and the last component of a multi-component burst or the centre of the fitted Gaussian for single-component bursts. Extended Data Figure~\ref{fig-extended:wait-times} displays the log-normal distribution of time separation between components of multi-component bursts. The Gaussian fit yields a characteristic separation of $\delta t=4.1^{+4.4}_{-2.1}~\mathrm{ms}$ between burst components which is a factor $2-10$ larger than that reported\cite{hessels_2019_apjl} for \rone.

\subsubsection*{Burst rates}\label{sec:results:burst-rates}

The distribution of bursts that we detected shows no evidence of a broken power law, which is why we fit a simple power law for the cumulative burst rate as function of spectral energy density, $R(>E_{\nu}) \propto E_{\nu}^{\gamma}$, using a least-squares method. For the fitting, we set a conservative $15\sigma$ completeness limit (assuming a canonical burst width of $1~\mathrm{ms}$) of $10~\mathrm{Jy~ms}$ and $39~\mathrm{Jy~ms}$ for O8 and St, respectively, and we adopt a $20$\% uncertainty on the derived spectral energy density. To better estimate the uncertainties of the fitted parameters, we once fit the entire data set and we also run a bootstrapping algorithm where we randomly select $90$\% of the data points (without replacement) and perform the least-squares fit on this subset of data. We repeat the bootstrapping step $1000$ times and take the standard deviation of the fitted parameters as the additional bootstrapping error. Jointly fitting the O8 and St data yields a power-law index $\gamma=-0.48\pm0.11\pm0.03$. We also fit the contemporaneous FAST data\cite{xu_2022_natur} in a similar fashion: we first apply our definition of a unique burst to the FAST data by summing fluences of bursts that are separated by less than $100~\mathrm{ms}$. This affected a total of $148$ bursts detected by FAST. We then determine the threshold $E_{\nu}^{\rm min}$ below which the high-energy part of the FAST bursts no longer follows a simple power-law using the Python {\texttt powerlaw}\cite{Alstott_2014_PLoSO} package and fit the data where $E > E_{\nu}^{\rm min}=1.7\times10^{30}\mathrm{~erg~Hz}^{-1}$ in the same way as described above to find $\gamma_{\mathrm{FAST}} = -2.254\pm0.019\pm0.114$, even steeper than the reported\cite{xu_2022_natur} $\gamma_2=-1.5\pm0.1\pm0.1$. \citet{xu_2022_natur} found the break point between the low- and high-energy range of their burst distribution to be $\mathrm{\sim5.9\times10^{29}~erg~Hz^{-1}}$ (using their median bandwidth of $185~\mathrm{MHz}$ to convert from energy to spectral energy density), roughly a factor $3$ lower than our adopted $\mathrm{E_{\nu}^{min}}$. Using the break point from \citet{xu_2022_natur} as the limiting energy for the fit, we find $\gamma_\mathrm{{FAST}} = -1.947\pm0.011\pm0.063$, which agrees with $\gamma_2$ within $2\sigma$.

For completeness, we also perform the fitting of the FAST data without summing components that are separated by less than $100~\mathrm{ms}$. This yields values for $\gamma_\mathrm{{FAST}}$ of $-2.318\pm0.018\pm0.116$ (when determining the power-law break point via {\texttt powerlaw}) and $-1.997\pm0.011\pm0.067$ (using the published break point\cite{xu_2022_natur}) which agree within the uncertainties with the values found when applying our definition of a unique burst.

The bursts and the associated cumulative burst rates that we detected with O8, Wb, and St after MJD~$59602$ are shown in Figure~\ref{fig:burstrates} (right). We fit the data from Wb and St in the same way as described above, however using a $15\sigma$ completeness limit of $14~\mathrm{Jy~ms}$ for both telescopes (because of the upgrade at St discussed above). We did not fit the bursts detected at O8 separately because of the low number of detections. The burst rates we find are $\mathrm{\gamma_{Wb}=-1.43\pm0.09\pm0.41}$ and $\mathrm{\gamma_{St}=-0.85\pm0.05\pm0.09}$ for the Wb and St bursts, respectively.

\subsubsection*{Implications of non-detections at P- and C-band}
As listed in Table~\ref{tab:coverage}, we spent about $650$~hr on \rsixseven with Westerbork at P-band, and another $244$~hr with Toru\'n at C-band. If we consider the time windows with enhanced activity only, i.e. MJD~$59305-59363$ and MJD~$59602-59641$ (Figure~\ref{fig:observations}), these numbers reduce to $274~\mathrm{hr}$ and $36~\mathrm{hr}$ for P-band and to $115~\mathrm{hr}$ and $86~\mathrm{hr}$ for C-band observations, respectively. During the same two time ranges, we spent a total of $513~\mathrm{hr}$ and $351~\mathrm{hr}$ of non-overlapping time on source at L-band. The completeness thresholds at P- and C-band are $\mathrm{\sim91~Jy~ms}$ and $\mathrm{\sim5~Jy~ms}$, respectively. In Extended Data Table~\ref{tab:p-c-band-bursts} we list the number of bursts that we detected at L-band above a given fluence threshold; once assuming a flat spectrum for the bursts and once assuming a steep spectrum with $\alpha=-1.5$. Given the number of bursts detected at L-band and the time observed at each band, we also list the number of bursts that we would have expected to detect at P- and C-band, assuming that activity rates are the same across all observed frequencies. The expected number of bursts ranges between $0.6$ and $7.6$ bursts, depending on observing band and epoch considered. The fact that we did not detect a single burst outside of L-band clearly shows that the bursts are narrow-band in nature\cite{majid_2020_apjl, gourdji_2019_apjl} and that activity rates and potentially even activity windows are frequency dependent\cite{pleunis_2021_apjl, pastormarazuela_2021_natur}. Alternatively, besides a spectral index, the emission process might be such that the amplitude of the bursts is modulated as a function of frequency, peaking in the range of $1200-1700~\mathrm{MHz}$ during our observations. Whatever the modulation mechanism might be (intrinsic to the source or a propagation effect such as refractive scintillation), it has to be time variable as the source was first detected by CHIME/FRB in the band between $400-800~\mathrm{MHz}$\cite{chime_2021_atel}.

\subsubsection*{The notch in burst B08}

We note a curious feature in burst B08, which is highlighted in Extended Data Figure~\ref{fig-extended:the-notch} for B08-tr (this can also be seen in Figure~\ref{fig-extended:2bit-demo} which shows the same burst detected at O8, B08-o8). We refer to this sudden, sharp dip in flux density as a `notch'. The B08 notch, lasting $\sim 0.5~\mathrm{ms}$, is somewhat reminiscent of the notches seen in the average profiles of a few very bright radio pulsars\cite{McLaughlin_2004_MNRAS}, which also happen to have exceptionally wide pulse profiles that cover a large range of rotational phases. However, pulsar notches are seen to be double notches (two closely spaced dips), whereas the limited S/N of B08 makes it impossible to see more subtle features in the notch profile. This limits the degree to which we can compare the two phenomena. 

In the case of pulsars, the notches have been modeled as coming from an absorbing region in the outer magnetosphere that co-rotates with the neutron star\cite{Wright_2004_MNRAS}. This model requires that the pulsar has both emitting and absorbing regions at heights that are a significant fraction of the light cylinder radius. In another model\cite{Dyks_2005_ApJ}, the pulsar itself is the absorber and the emission is a combination of inward and outward travelling components. 

FRBs often show quite well defined sub-bursts with cessation of emission between these components\cite{hessels_2019_apjl}; this is also clearly visible in B08 (Extended Data Figure~\ref{fig-extended:the-notch}, left panel). However, the notch we detect in B08 appears qualitatively different: it is a sudden sharp dip on the shoulder of a bright component, much like what is seen in pulsar notches (Extended Data Figure~\ref{fig-extended:the-notch}, right panel). We speculate that notches may be generally visible in the profiles of high-S/N, wide FRBs observed at high time resolution, and that this effect could plausibly probe magnetospheric physics, assuming such an origin for FRB emission. Furthermore, if notches can be detected in multiple bursts from a single repeater, then these could provide a stable marker of rotational phase that could allow the spin rate of the neutron star to be determined. This assumes that the absorber is locked in rotational phase with respect to the neutron star. For pulsars, the notches indeed occur at stable rotational phases and this is explainable in the models\cite{Wright_2004_MNRAS, Dyks_2005_ApJ} mentioned above. FRB notches would also potentially be excellent reference points for determining accurate DMs; as can be seen in Extended Data Figure~\ref{fig-extended:the-notch}, the notch occurs at the same time across the observed radio frequency band for the DM we have chosen to dedisperse all bursts in the paper. 

\section*{Additional information}
Correspondence and requests for material should be addressed to F.K.
\section*{Acknowledgements}
We would like to thank the directors and staff at the various participating telescopes for allowing us to use their facilities and for helping to run the observations.
We would like to express our gratitude to Willem van Straten for modifying the DSPSR software package to fit our needs and for helping is with the SPC-algorithm.
F.K. acknowledges support from Onsala Space Observatory for  the  provisioning of its facilities/observational support. The Onsala Space Observatory national research infrastrcuture is funded through Swedish Research Council grant No 2017-00648.
Research by the AstroFlash group at University of Amsterdam, ASTRON and JIVE is supported in part by an NWO Vici grant (PI Hessels; VI.C.192.045).
This work is based in part on observations carried out using the 32-m radio telescope operated by the Institute of Astronomy of the Nicolaus Copernicus University in Toru\'n (Poland) and supported by a Polish Ministry of Science and Higher Education SpUB grant.
This work makes use of data from the Westerbork Synthesis Radio Telescope owned by ASTRON. ASTRON, the Netherlands Institute for Radio Astronomy, is an institute of the Dutch Scientific Research Council NWO (Nederlandse Oranisatie voor Wetenschappelijk Onderzoek).

\section*{Author contributions}
F.K. wrote and ran the search pipeline, led the observations at Onsala, interpreted the data and led the paper writing. O.S.O.B led the observations at Westerbork, created Figs. 1--4 and Extended Data Figs. 1--3. W.H. led the observations at Stockert, searched the Stockert data for bursts and described the observations. M.P.G. led the observations at Toru\'n and searched the data for bursts. J.W.T.H. interpreted the data scientifically, supervised student work, and wrote parts of the manuscript. W.L. interpreted the data scientifically and commented on the manuscript. M.P.S., P.C., K.N., and R.v.R provided comments on the manuscript. J.Y. supported the observations at Onsala. R.B. supported the observations at Westerbork. W.P. and P.W. supported the observations at Toru\'n.

\section*{Competing interests}
The authors declare no competing interests.

\begin{table*} 
\caption{\label{tab:coverage}Observational setup.}
\begin{tabular}{lccccccc}
\hline
\hline
Station$\mathrm{^{a}}$  & Band$\mathrm{^{b}}$ & Frequency & Bandwidth$\mathrm{^{c}}$ & Bandwidth per & SEFD$\mathrm{^{d}}$ & Completeness$\mathrm{^{e}}$ & Time observed \\
& & [MHz] & [MHz] & subband [MHz] & [Jy] & [Jy~ms] & [hr] \\
\hline
Wb  & P                 & 300--364          &60     & 8           & 2100  & 91     & 649.51\\
Wb  & L$_{\rm Wb-1}$    & 1259--1387        &100    & 16          & 420   & 14     & 383.1 \\
Wb  & L$_{\rm Wb-2}$    & 1382--1510        &100    & 16          & 420   & 14    & 129.5 \\
St  & L$_{\rm St}$    & 1332.5--1430.5    &98     & 98         & 1100/385 & 39/14     & 1431.7 \\
O8  & L$_{\rm O8-1}$    & 1360--1488        &100    & 16          & 310   & 10     & 415.7 \\
O8  & L$_{\rm O8-2}$    & 1460--1588        &100    & 16          & 310   & 10     & 85.7 \\
O8  & L$_{\rm O8-3}$    & 1360--1616        &200    & 16          & 310   & 7     & 9.8 \\
O8  & L$_{\rm O8-4}$    & 1202--1714        &350    & 32          & 310   & 5     & 40.9 \\
Tr  & L$_{\rm Tr-1}$    & 1290--1546         &200   & 32          & 350   & 8     & 116.9 \\
Tr  & L$_{\rm Tr-2}$    & 1340--1596         &200   & 32          & 350   & 8     & 22.4 \\
Tr  & L$_{\rm Tr-3}$    & 1350--1478         &100   & 16          & 350   & 12     & 15.6 \\
Tr  & C$_{\rm Tr-1}$    & 4550--4806         &240   & 32          & 220   & 5     & 227.7 \\
Tr  & C$_{\rm Tr-2}$    & 4600--4728         &120   & 32          & 220   & 7     & 16.4 \\
\hline
\multicolumn{7}{l}{Total telescope time/total time on source [hr]$\mathrm{^{f}}$} & 3545/2281 \\
\hline

\multicolumn{8}{l}{$\mathrm{^{a}}$ Wb: Westerbork RT1, O8: Onsala $\mathrm{25-m}$, Tr: Toru\'n, St: Stockert} \\
\multicolumn{8}{l}{$\mathrm{^{c}}$ Effective bandwidth accounting for RFI and band edges.} \\
\multicolumn{8}{l}{$\mathrm{^{d}}$ From the \href{http://old.evlbi.org/user_guide/EVNstatus.txt}{EVN status page} (with the exception of St, where the two values indicate the pre- and post-upgrade sensitivity).} \\
\multicolumn{8}{l}{$\mathrm{^{e}}$ Assuming a $15\sigma$ detection threshold and a pulse width of $1~\mathrm{ms}$.} \\
\multicolumn{8}{l}{$\mathrm{^{f}}$ Total time on source accounts for overlap between the participating stations.} \\
\end{tabular}
\end{table*}

\begin{table*} 
\caption{\label{tab:numbers-per-epoch}Observing hours and number of bursts detected per active period and dish}
\begin{tabular}{ccccc}
\hline
\hline
        & \multicolumn{2}{c}{Epoch 1$\mathrm{^{a}}$} & \multicolumn{2}{c}{Epoch 2$\mathrm{^{b}}$}\\
Station & Hours$\mathrm{^{c}}$ & N$\mathrm{^{d}}$ & Hours$\mathrm{^{c}}$ & N$\mathrm{^{d}}$ \\
\hline
Wb      &  9   & 0 & 199 & 24\\
O8      & 257  & 8 & 58  & 6 \\
Tr      & 22   & 1 & 16  & 1 \\
St      & 390  & 3 & 178 & 13 \\
\hline
\multicolumn{5}{l}{$\mathrm{^{a}}$ MJD~$59305-59363$} \\
\multicolumn{5}{l}{$\mathrm{^{b}}$ MJD~$59602-59642$} \\
\multicolumn{5}{l}{$\mathrm{^{c}}$ Hours spent on source at L-band} \\
\multicolumn{5}{l}{$\mathrm{^{d}}$ Number of bursts detected} \\
\end{tabular}
\end{table*}

\clearpage
\onecolumn

\section*{Extended Data}
\setcounter{figure}{0}
\captionsetup[figure]{name={\bf Extended Data Figure}}

\setcounter{table}{0}
\captionsetup[table]{name={\bf Extended Data Table}}

\pagestyle{empty}
\begin{figure*}[htb!]
    \centering
    \includegraphics[width=\textwidth]{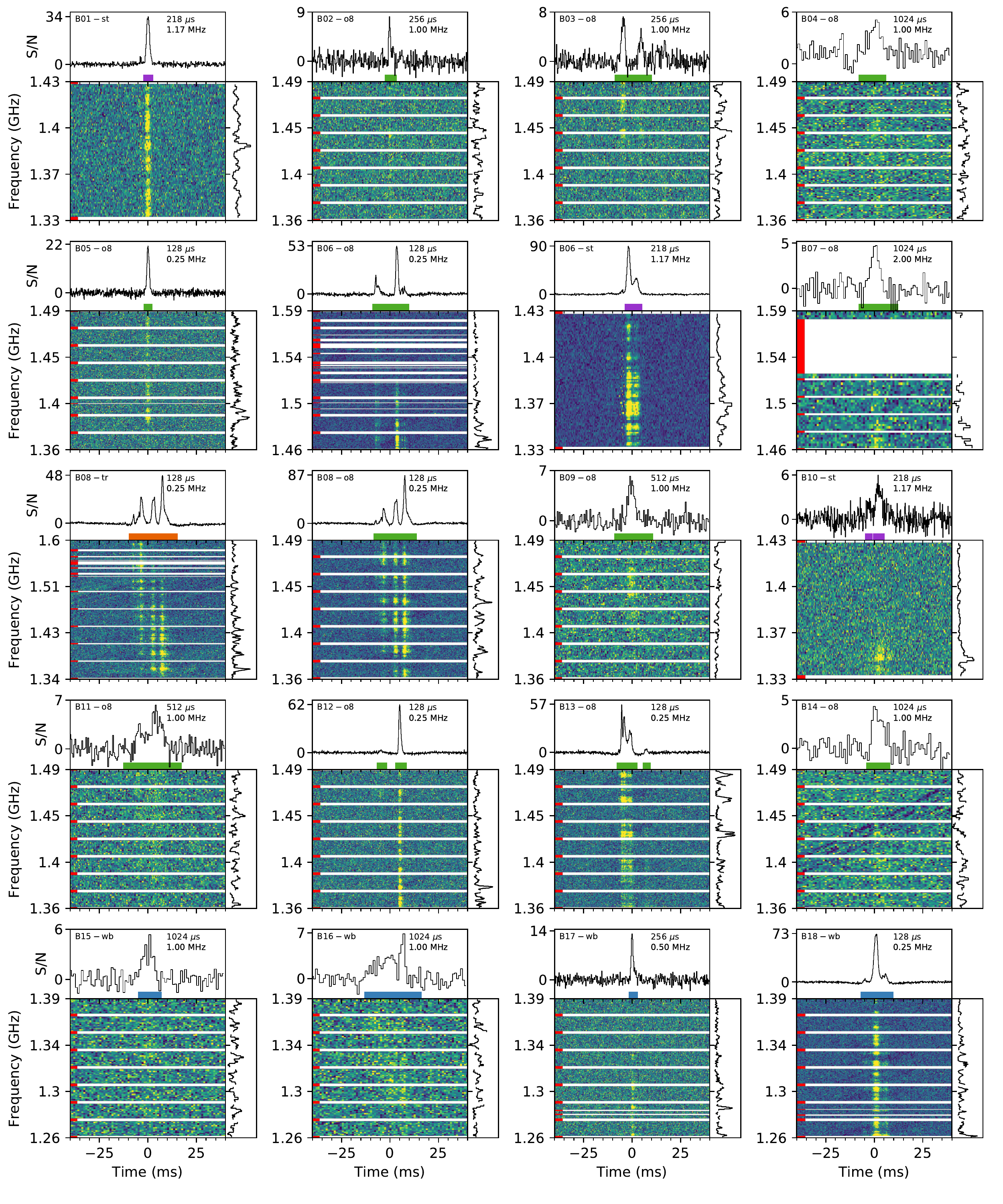}
    \caption{Plot of all bursts detected in this campaign. See Figure~\ref{fig:subset-of-figs} for a full description of the sub-panels.}
    \label{fig-extended:familyplot}
\end{figure*}
\newpage
\begin{Contfigure*}[htb!]
    \centering
    \includegraphics[width=\textwidth]{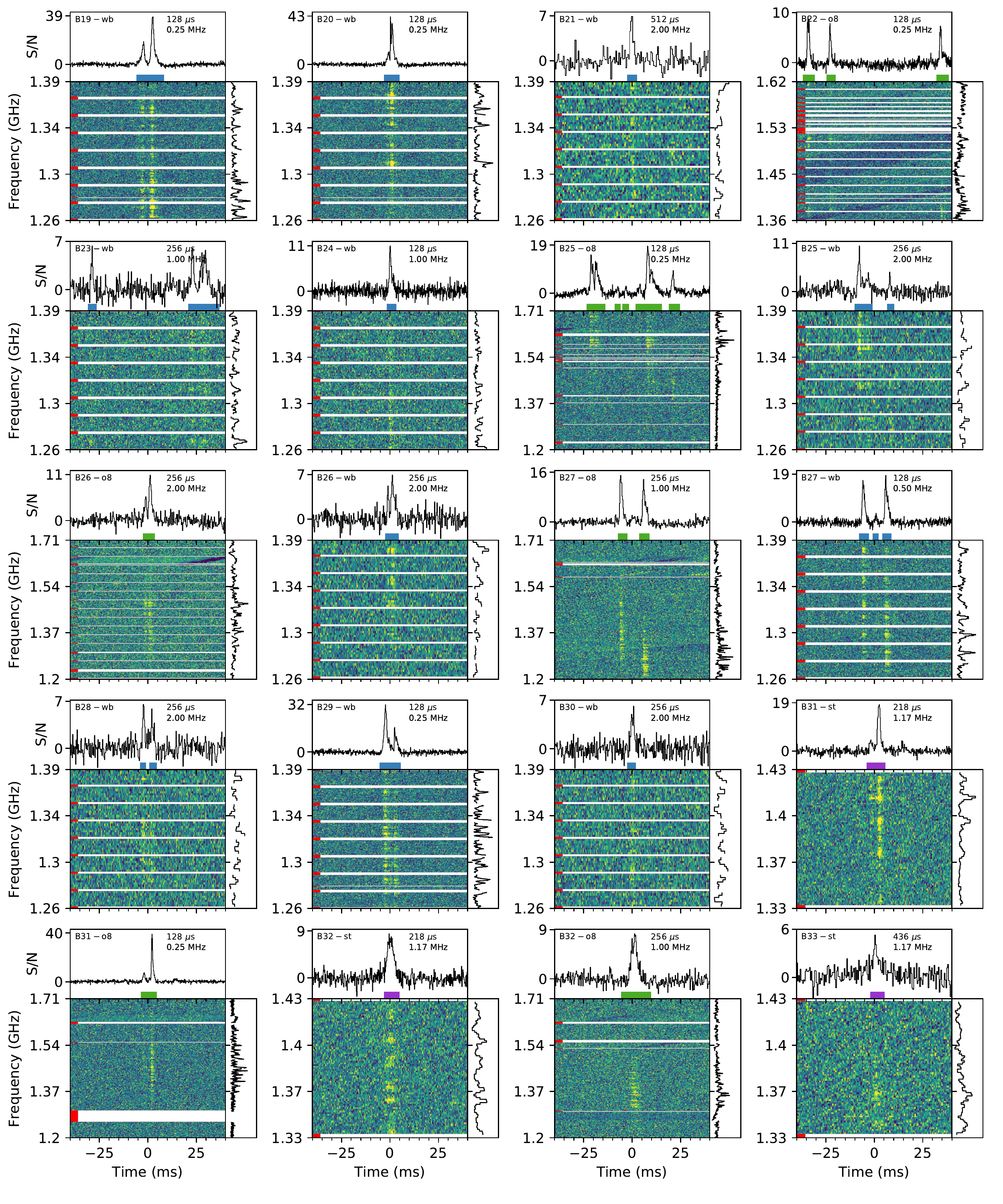}
    \caption{}
\end{Contfigure*}
\newpage
\begin{Contfigure*}[htb!]
    \centering
    \includegraphics[width=\textwidth]{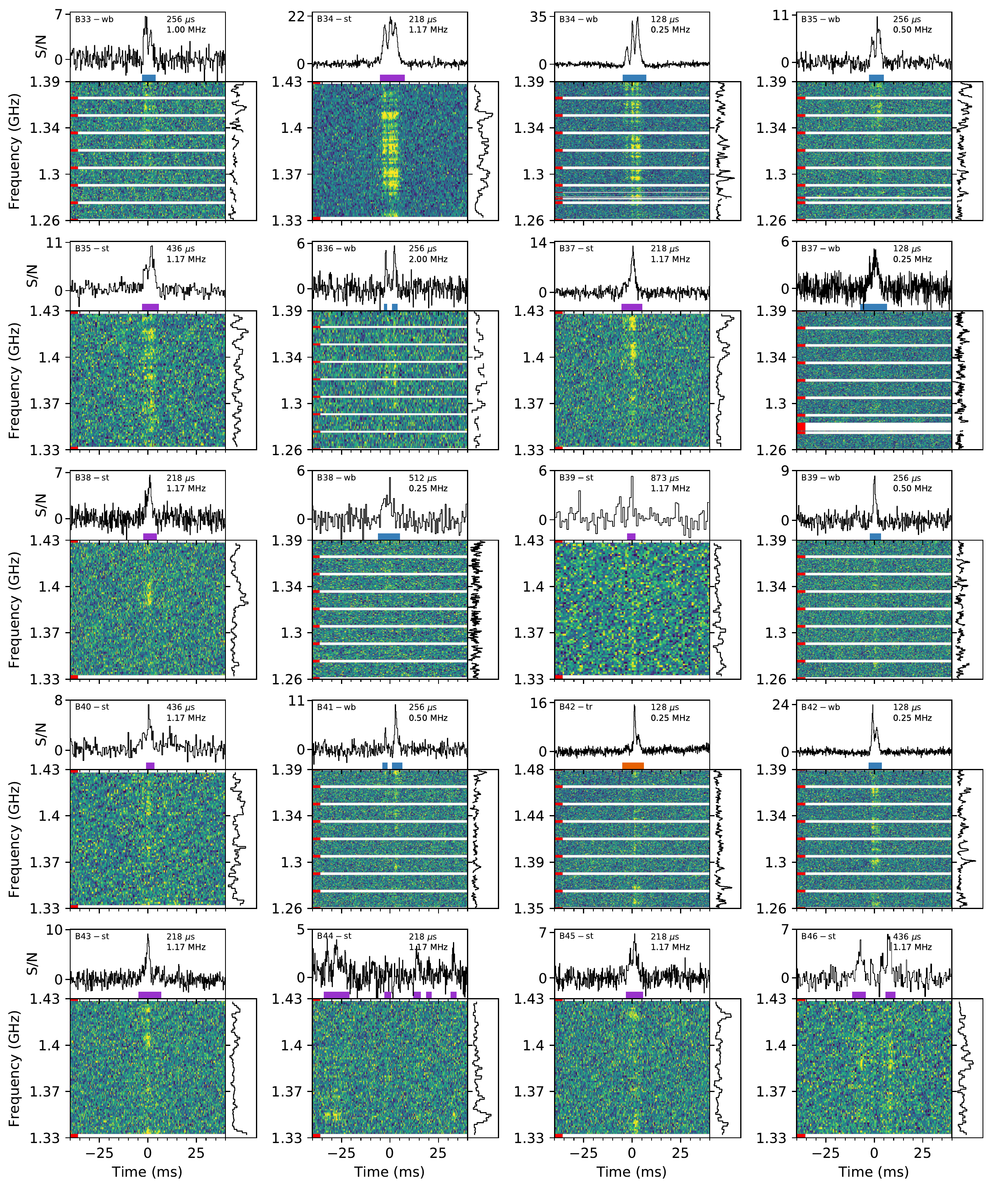}
    \caption{}
\end{Contfigure*}

\begin{figure*}
    \centering
    \includegraphics[width=0.3\textwidth]{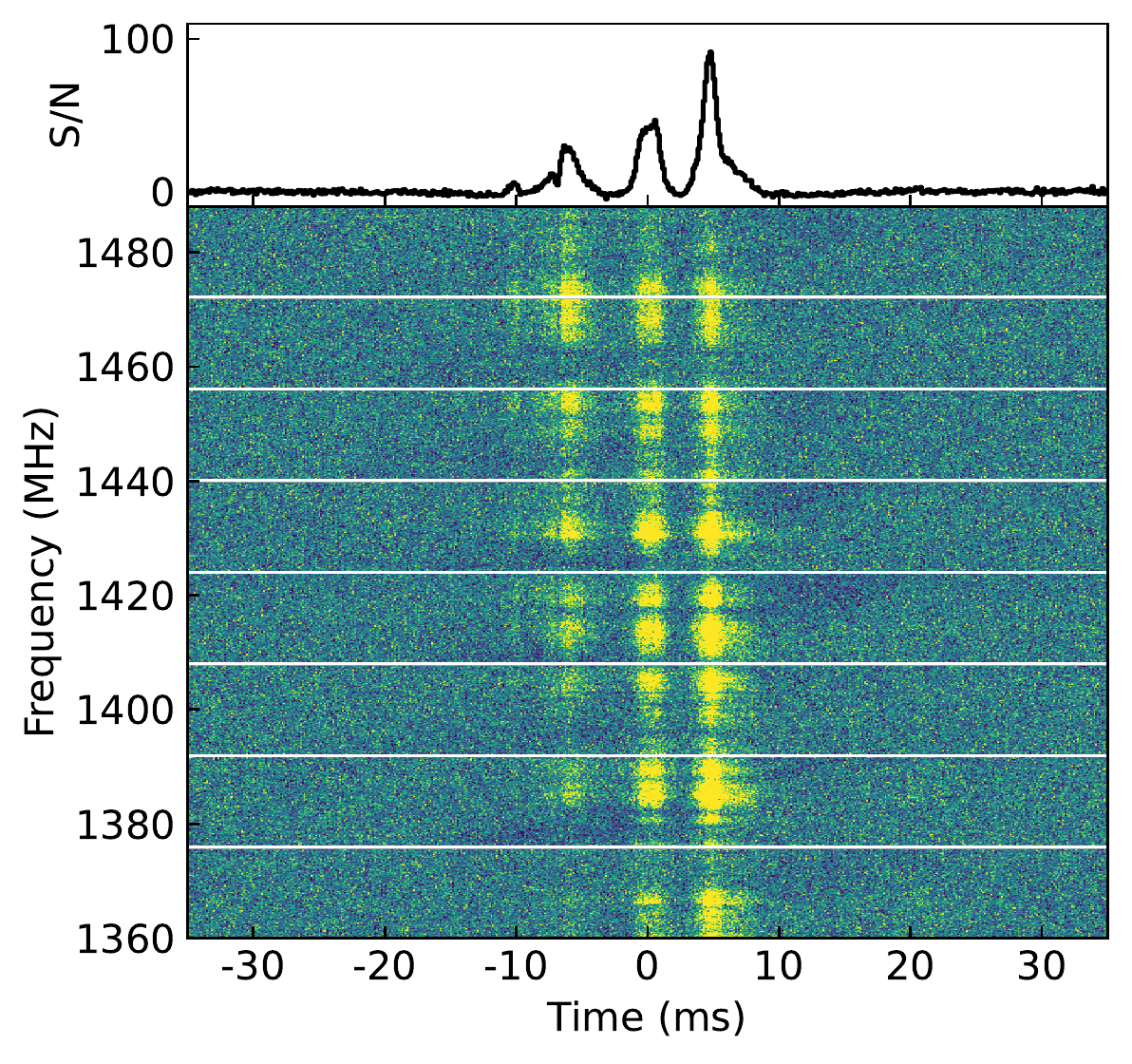}
    \includegraphics[width=0.3\textwidth]{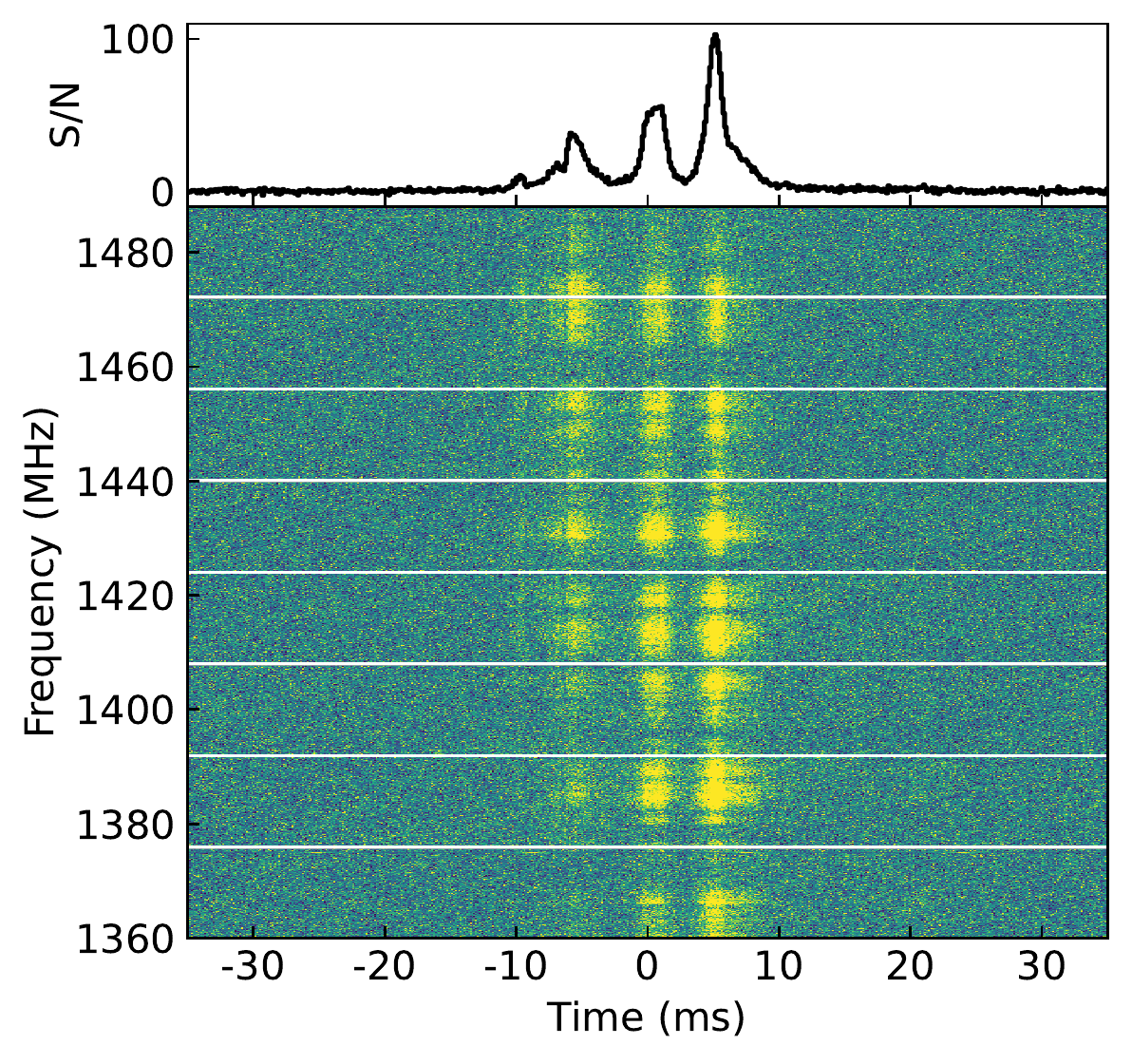}
    \includegraphics[width=0.3\textwidth]{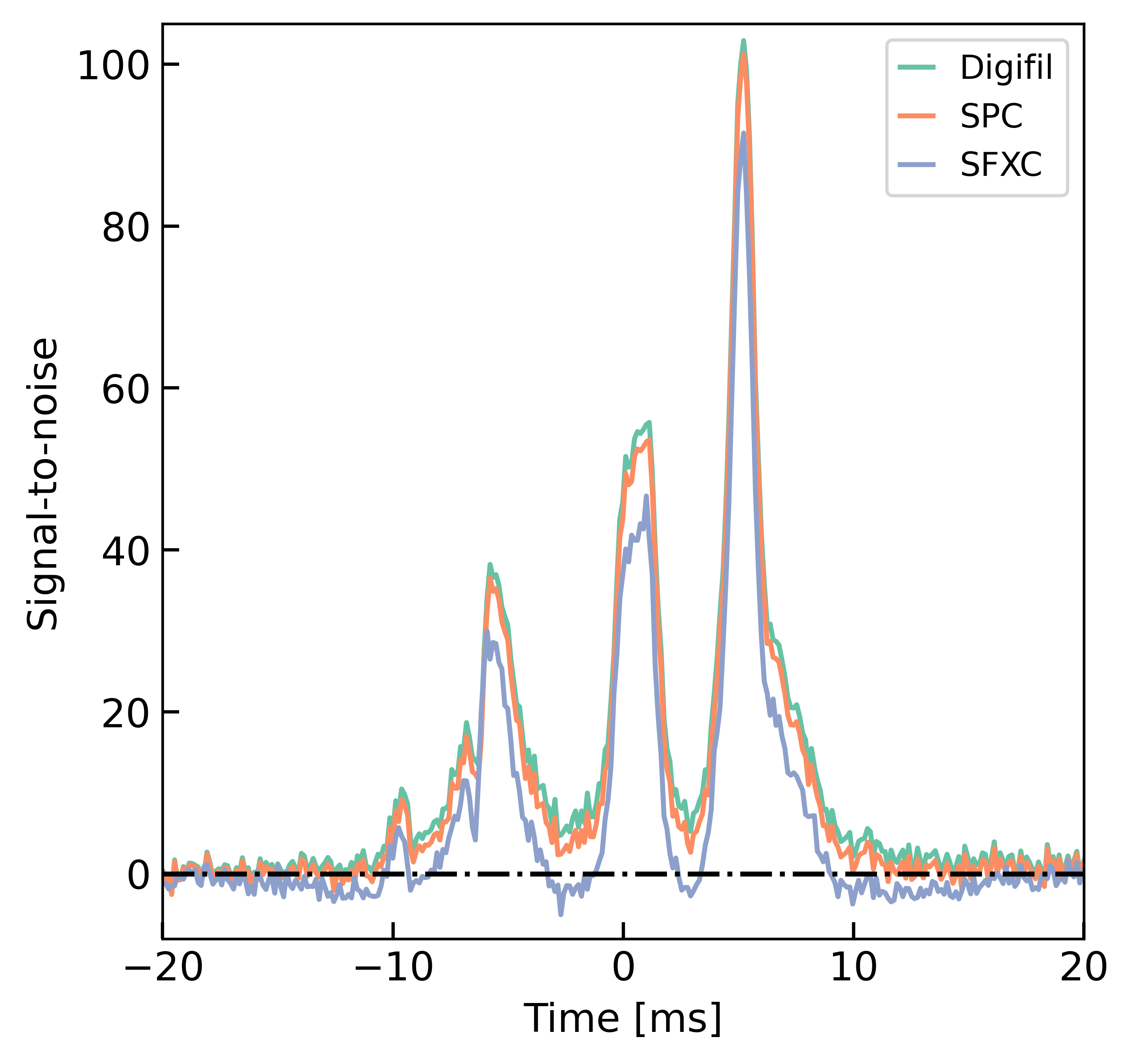}
    \caption{Illustration of the effect of $2$-bit sampling on the total intensity data, here shown for burst B08-o8. Left: Processed with \sfxc; the top panel shows the frequency-averaged time series while the bottom panel shows the dedispersed dynamic spectrum. The response of the system to the brightest scintillation peaks is visible as dark `depressions' across the affected subbands (the edges of subbands are indicated by horizontal white lines). Middle: The same burst processed with \digifil. The $2$-bit correction removes the depression but, at the same time, overcompensates for this instrumental effect. Right: Overlay of the time series as processed with \sfxc, \digifil and also further processed with \psrchive's SPC algorithm. While the \sfxc-generated profile dips below the baseline between components, the SPC-processed version remains below the \digifil time series while not reaching the zero-baseline in between components.}
    \label{fig-extended:2bit-demo}
\end{figure*}
\begin{figure}
    \centering
    \includegraphics[width=0.9\columnwidth]{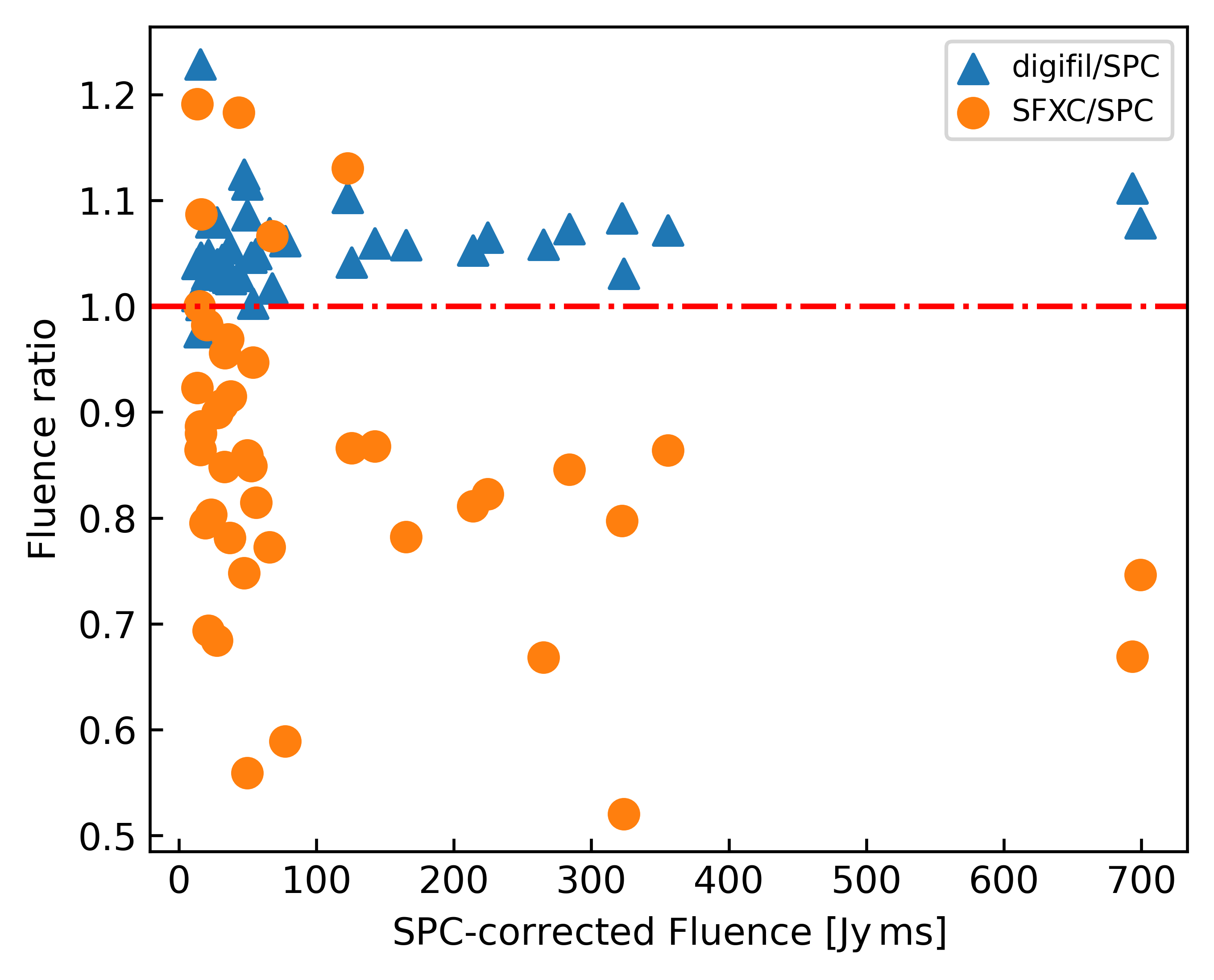}
    \caption{Relative fluences computed from the \sfxc- and \digifil-generated filterbanks, scaled by the SPC-corrected fluences (see text for details).}
    \label{fig-extended:fluence-ratio}
\end{figure}
\begin{figure}
    \centering
    \includegraphics[width=0.9\columnwidth]{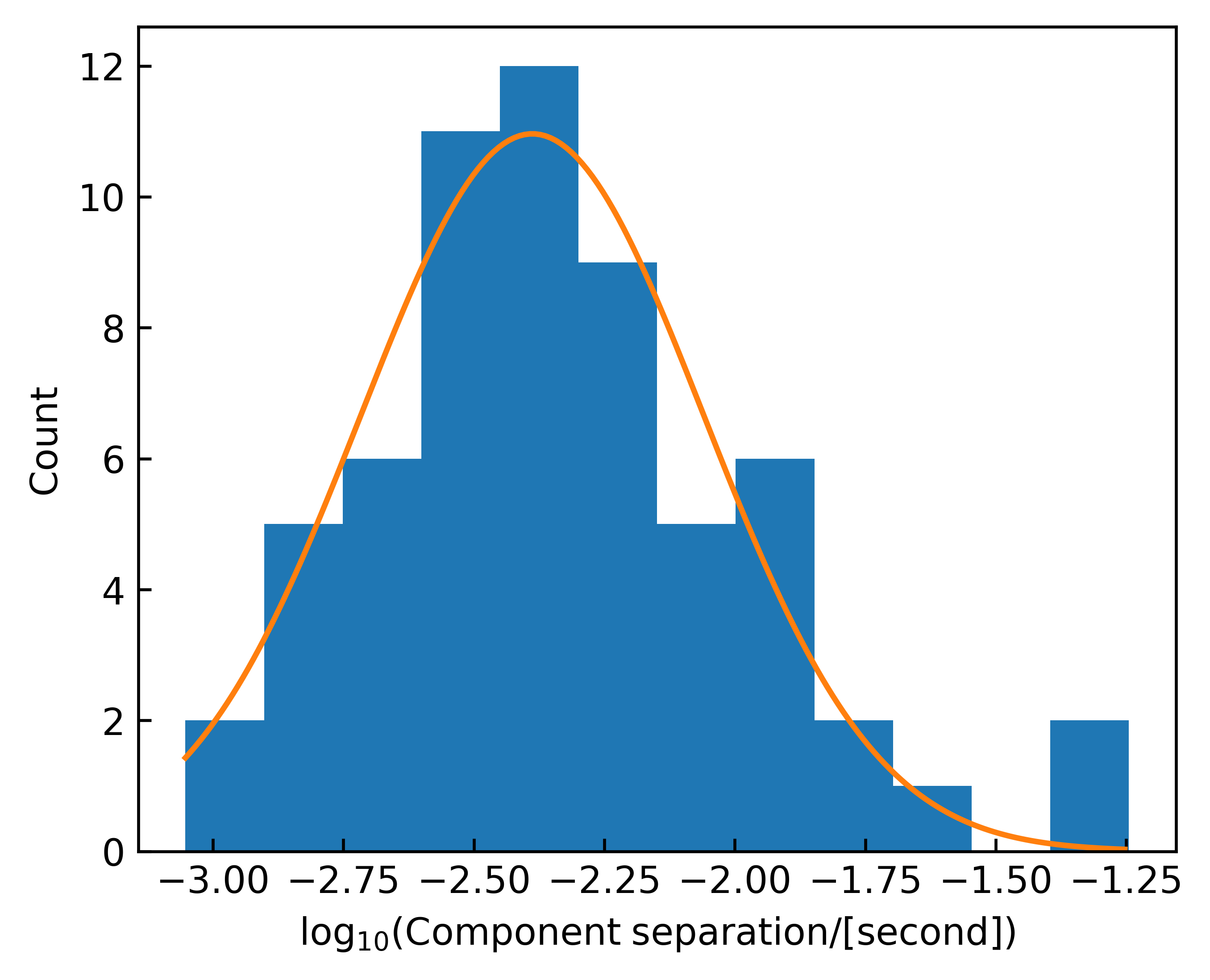}
    \caption{Logarithmic distribution of wait times between burst components. The orange line is a log-normal fit that yields a characteristic wait time of $\delta t=4.1^{+4.4}_{-2.1}$~ms.}
    \label{fig-extended:wait-times}
\end{figure}

\begin{figure}
    \centering
    \includegraphics[width=0.9\columnwidth]{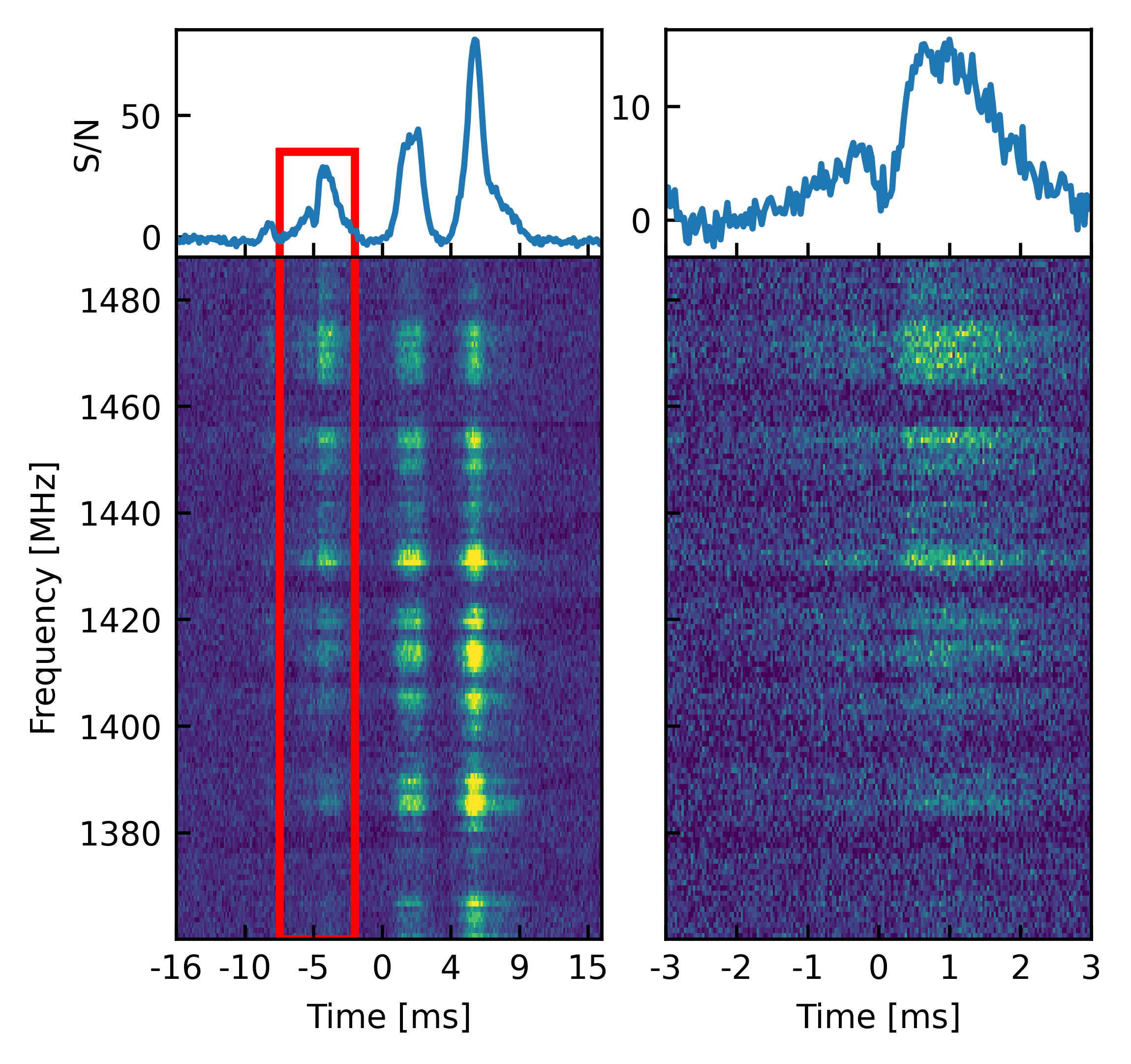}
    \caption{Coherently dedispersed burst B08-tr. Left: The full burst at a time and frequency resolution of $128~\mu$$\mathrm{s}$ and $1~\mathrm{MHz}$. The red box indicates the region that was zoomed in on on the right. Right: Zoom in on burst B08-tr highlighting the `notch'-like feature. The data are plotted at a time and frequency resolution of 32~$\mu$$\mathrm{s}$ and $1~\mathrm{MHz}$, respectively.}
    \label{fig-extended:the-notch}
\end{figure}
\begin{table*}
\caption{\label{tab:burst_properties}Burst properties.}
\footnotesize 
\begin{tabular}{lcccccccc}\hline
\hline
Burst ID & Station & {TOA$\mathrm{^{a}}$} & {Peak S/N$\mathrm{^{b}}$} & {Fluence$\mathrm{^{c}}$} & Number of & {Width$\mathrm{^{d}}$} & {Spectral Luminosity$\mathrm{^{e}}$} & {$\nu_{s}\mathrm{^{f}}$}\\ 
  &  & {[MJD]} &  & {[Jy~ms]} & components & {[ms]} & {[$\mathrm{10^{32}\,erg\,s^{-1}\,Hz^{-1}}$]} & {[MHz]}\\ 
\hline 
B01-st & St & 59319.513506069* & 45.7   & $383.1\pm76.6$ & 1   &    5.24   & $148.83\pm29.77$  & $0.4\pm0.3$ \\ 
B02-o8 & O8 & 59326.642775092 & 7.6   & $13.3\pm2.7$ & 1   &    6.66   & $4.07\pm0.81$  & $0.9\pm0.8$ \\ 
B03-o8 & O8 & 59329.517962388 & 7.6   & $49.6\pm9.9$ & 3   &    28.93   & $3.49\pm0.70$  & $0.4\pm0.3$ \\ 
B04-o8 & O8 & 59337.453941128* & 4.5   & $28.8\pm5.8$ & 1   &    14.85   & $3.95\pm0.79$  & $0.9\pm1.7$ \\ 
B05-o8 & O8 & 59337.797038104 & 28.5   & $53.8\pm10.8$ & 1   &    6.66   & $16.48\pm3.30$  & $0.4\pm0.3$ \\ 
B06-o8 & O8 & 59349.567736580 & 76.9   & $265.2\pm53.0$ & 4   &    19.46   & $27.76\pm5.55$  & $0.4\pm0.1$ \\ 
B06-st & St & 59349.567736521* & 114.5   & $1593.6\pm318.7$ & 3   &    9.18   & $353.80\pm70.76$  & $0.2\pm0.1$ \\ 
B07-o8 & O8 & 59349.572426981 & 3.4   & $15.5\pm3.1$ & 1   &    27.39   & $1.15\pm0.23$  & $0.1\pm0.5$ \\ 
B08-tr & Tr & 59356.468854046 & 108.2   & $453.0\pm90.6$ & 4   &    23.04   & $40.05\pm8.01$  & $0.5\pm0.1$ \\ 
B08-o8 & O8 & 59356.468854047 & 137.4   & $693.4\pm138.7$ & 4   &    23.81   & $59.32\pm11.86$  & $0.2\pm0.2$ \\ 
B09-o8 & O8 & 59358.428680977 & 5.8   & $23.5\pm4.7$ & 1   &    15.36   & $3.11\pm0.62$  & $0.3\pm0.5$ \\ 
B10-st & St & 59362.433401658* & 9.0   & $278.2\pm55.6$ & 3   &    10.05   & $56.39\pm11.28$  & $0.2\pm0.2$ \\ 
B11-o8 & O8 & 59485.203789751 & 5.2   & $47.5\pm9.5$ & 2   &    29.18   & $3.31\pm0.66$  & $0.4\pm0.4$ \\ 
B12-o8 & O8 & 59485.238856691 & 88.6   & $165.1\pm33.0$ & 2   &    17.66   & $19.04\pm3.81$  & $0.3\pm0.2$ \\ 
B13-o8 & O8 & 59485.970826378 & 76.4   & $322.3\pm64.5$ & 5   &    18.69   & $35.13\pm7.03$  & $0.6\pm0.1$ \\ 
B14-o8 & O8 & 59485.983419756 & 3.5   & $15.8\pm3.2$ & 1   &    17.41   & $1.85\pm0.37$  & $0.6\pm0.4$ \\ 
B15-wb & Wb & 59602.992809140 & 4.6   & $27.7\pm5.5$ & 1   &    15.36   & $3.68\pm0.74$  & $0.3\pm0.4$ \\ 
B16-wb & Wb & 59603.754507174 & 6.2   & $49.8\pm10.0$ & 1   &    33.02   & $3.07\pm0.61$  & $3.8\pm4.2$ \\ 
B17-wb & Wb & 59603.799227784 & 14.0   & $33.1\pm6.6$ & 1   &    5.38   & $12.56\pm2.51$  & $0.2\pm0.2$ \\ 
B18-wb & Wb & 59605.835730583 & 122.1   & $699.2\pm139.8$ & 3   &    17.15   & $83.03\pm16.61$  & $0.3\pm0.1$ \\ 
B19-wb & Wb & 59608.845447891 & 57.5   & $284.0\pm56.8$ & 2   &    15.87   & $36.45\pm7.29$  & $0.3\pm0.2$ \\ 
B20-wb & Wb & 59608.888891277 & 52.9   & $213.7\pm42.7$ & 3   &    8.70   & $50.02\pm10.00$  & $0.5\pm0.1$ \\ 
B21-wb & Wb & 59609.758499853 & 6.1   & $16.0\pm3.2$ & 1   &    5.89   & $5.53\pm1.11$  & $0.7\pm0.5$ \\ 
B22-o8 & O8 & 59611.690224668 & 17.3   & $67.7\pm13.5$ & 3   &    75.01   & $1.84\pm0.37$  & $1.4\pm0.6$ \\ 
B23-wb & Wb & 59611.911786572 & 6.5   & $66.0\pm13.2$ & 3   &    68.35   & $1.97\pm0.39$  & $0.6\pm0.3$ \\ 
B24-wb & Wb & 59612.971443269 & 14.7   & $37.8\pm7.6$ & 1   &    4.86   & $15.82\pm3.16$  & $0.3\pm0.3$ \\ 
B25-o8 & O8 & 59613.913140698 & 30.2   & $122.8\pm24.6$ & 8   &    46.59   & $5.37\pm1.07$  & $0.3\pm0.1$ \\ 
B25-wb & Wb & 59613.913141220 & 10.8   & $56.3\pm11.3$ & 3   &    19.20   & $5.97\pm1.19$  & $0.1\pm0.1$ \\ 
B26-o8 & O8 & 59614.043244112 & 10.5   & $14.9\pm3.0$ & 2   &    6.14   & $4.93\pm0.99$  & $0.4\pm0.2$ \\ 
B26-wb & Wb & 59614.043244218 & 7.4   & $36.9\pm7.4$ & 2   &    7.68   & $9.79\pm1.96$  & $0.7\pm0.3$ \\ 
B27-o8 & O8 & 59614.048814839 & 23.4   & $77.2\pm15.4$ & 2   &    16.90   & $9.30\pm1.86$  & $0.5\pm0.3$ \\ 
B27-wb & Wb & 59614.048814942 & 24.6   & $142.5\pm28.5$ & 3   &    16.90   & $17.18\pm3.44$  & $0.5\pm0.2$ \\ 
B28-wb & Wb & 59615.749713619 & 6.9   & $28.0\pm5.6$ & 2   &    8.45   & $6.76\pm1.35$  & $0.1\pm0.2$ \\ 
B29-wb & Wb & 59615.845856436 & 41.4   & $224.6\pm44.9$ & 2   &    11.78   & $38.86\pm7.77$  & $0.2\pm0.1$ \\ 
B30-wb & Wb & 59616.969495614 & 5.4   & $19.3\pm3.9$ & 1   &    6.40   & $6.14\pm1.23$  & $0.2\pm0.3$ \\ 
B31-st & St & 59618.924807243* & 21.3   & $64.5\pm12.9$ & 2   &    9.39   & $13.99\pm2.80$  & $0.4\pm0.3$ \\ 
B31-o8 & O8 & 59618.924807260 & 55.1   & $51.1\pm10.2$ & 2   &    8.19   & $12.72\pm2.54$  & $0.4\pm0.2$ \\ 
B32-st & St & 59618.962324541* & 9.2   & $57.4\pm11.5$ & 2   &    7.86   & $14.88\pm2.98$  & $0.3\pm0.2$ \\ 
B32-o8 & O8 & 59618.962324547 & 9.8   & $21.6\pm4.3$ & 1   &    7.94   & $5.54\pm1.11$  & $0.4\pm0.2$ \\ 
B33-st & St & 59619.730729859* & 5.6   & $26.4\pm5.3$ & 1   &    7.43   & $7.24\pm1.45$  & $0.6\pm0.6$ \\ 
B33-wb & Wb & 59619.730730266 & 7.7   & $33.6\pm6.7$ & 2   &    7.42   & $9.22\pm1.84$  & $0.2\pm0.2$ \\ 
B34-st & St & 59619.773110130* & 27.5   & $250.7\pm50.1$ & 3   &    12.89   & $39.62\pm7.92$  & $0.6\pm0.2$ \\ 
B34-wb & Wb & 59619.773110529 & 52.0   & $355.6\pm71.1$ & 3   &    14.85   & $48.78\pm9.76$  & $0.5\pm0.2$ \\ 
B35-wb & Wb & 59620.855108645 & 9.4   & $52.4\pm10.5$ & 2   &    7.68   & $13.91\pm2.78$  & $0.3\pm0.2$ \\ 
B35-st & St & 59620.856761425* & 7.8   & $52.0\pm10.4$ & 2   &    8.96   & $11.83\pm2.37$  & $0.2\pm0.2$ \\ 
B36-wb & Wb & 59620.913032971 & 6.6   & $20.4\pm4.1$ & 2   &    7.42   & $5.59\pm1.12$  & --$\mathrm{^{g}}$ \\ 
B37-st & St & 59622.759765600* & 12.1   & $50.0\pm10.0$ & 1   &    10.70   & $9.51\pm1.90$  & $1.2\pm0.8$ \\ 
B37-wb & Wb & 59622.759765813 & 5.9   & $35.8\pm7.2$ & 1   &    10.75   & $6.79\pm1.36$  & $0.5\pm0.2$ \\ 
B38-st & St & 59622.795156022* & 6.7   & $26.2\pm5.2$ & 2   &    6.99   & $7.62\pm1.52$  & $3.8\pm4.1$ \\ 
B38-wb & Wb & 59622.795156240 & 3.7   & $13.4\pm2.7$ & 1   &    8.70   & $3.15\pm0.63$  & $0.5\pm0.4$ \\ 
B39-st & St & 59622.962155172* & 3.4   & $11.0\pm2.2$ & 1   &    4.37   & $5.13\pm1.03$  & --$\mathrm{^{g}}$ \\ 
B39-wb & Wb & 59622.962155576 & 7.3   & $16.2\pm3.2$ & 1   &    6.14   & $5.37\pm1.07$  & $0.4\pm0.3$ \\ 
B40-st & St & 59624.668435558* & 5.3   & $16.6\pm3.3$ & 1   &    4.37   & $7.73\pm1.55$  & $0.2\pm0.2$ \\ 
B41-wb & Wb & 59633.774378172 & 9.0   & $31.6\pm6.3$ & 2   &    9.22   & $6.99\pm1.40$  & $0.4\pm0.2$ \\ 
B42-tr & Tr & 59633.927686106 & 25.1   & $61.0\pm12.2$ & 3   &    11.01   & $11.30\pm2.26$  & $0.4\pm0.2$ \\ 
B42-wb & Wb & 59633.927686140 & 32.0   & $125.5\pm25.1$ & 2   &    7.68   & $33.28\pm6.66$  & $0.6\pm0.1$ \\ 
B43-st & St & 59635.821821805* & 9.6   & $45.0\pm9.0$ & 1   &    11.58   & $7.92\pm1.58$  & $0.6\pm0.3$ \\ 
B44-st & St & 59636.618137128* & 4.9   & $71.5\pm14.3$ & 6   &    68.38   & $2.13\pm0.43$  & $0.2\pm0.2$ \\ 
B45-st & St & 59637.952097942* & 6.6   & $36.1\pm7.2$ & 2   &    8.74   & $8.42\pm1.68$  & $1.0\pm0.4$ \\ 
B46-st & St & 59639.813991726* & 5.8   & $38.6\pm7.7$ & 2   &    21.85   & $3.60\pm0.72$  & $0.6\pm1.0$ \\ 
\hline
\multicolumn{9}{l}{$\mathrm{^{a}}$ Time of arrival at the solar system barycenter at infinite frequency in TDB (using a DM of $410.8$~\dmunit, a dispersion measure} \\ 
\multicolumn{9}{l}{\hspace{0.25cm}constant of $1/0.000241$ $\mathrm{GHz^2~cm^3~pc^{-1}~\mu s}$ and as (J2000) position RA = $05\!:\!08\!:\!03.5$, Dec = $+26\!:\!03\!:\!37.8$; for times} \\ 
\multicolumn{9}{l}{\hspace{0.25cm}marked with a *, a DM constant of $4.14880568679703$ $\mathrm{GHz^2\,cm^3\,pc^{-1}\,ms}$ was used).} \\ 
\multicolumn{9}{l}{\hspace{0.25cm}For multi-component busts, the TOA is defined as the middle between the peak of the first and the last component.} \\ 
\multicolumn{9}{l}{$\mathrm{^{b}}$ The peak S/N of the brightest component.} \\ 
\multicolumn{9}{l}{$\mathrm{^{c}}$ Computed as the sum over the measured fluence of each component. We assume a conservative error of $20$\% for all bursts, dominated by the} \\ 
\multicolumn{9}{l}{\hspace{0.25cm}uncertainty of the SEFD.} \\ 
\multicolumn{9}{l}{$\mathrm{^{d}}$ Manually determined time span between start of first and end of last component.} \\ 
\multicolumn{9}{l}{$\mathrm{^{e}}$ Computed using $D_L=453~\mathrm{Mpc}$, $z=0.098$ and the listed width.} \\ 
\multicolumn{9}{l}{$\mathrm{^{f}}$ Weighted average over the measured scintillation bandwidth per component.} \\ 
\multicolumn{9}{l}{$\mathrm{^{g}}$ No measurement was possible due to low S/N.} \\ 
\hline
\end{tabular}
\normalsize
\end{table*}

\begin{table*}
\caption{\label{tab:p-c-band-bursts}Expected burst numbers at P- and C-band.}
\begin{tabular}{ccccc|ccc}
\hline
\hline
 &          & \multicolumn{3}{c}{MJD range 1} & \multicolumn{3}{c}{MJD range 2} \\
Spectral    & Threshold$\mathrm{^{a}}$        & \multicolumn{3}{c}{59305--59363} & \multicolumn{3}{c}{59602--59641} \\
\cline{3-8}
index       & [Jy~ms]                          &  L$\mathrm{^{b}}$  & P$\mathrm{^{c}}$  & C$\mathrm{^{c}}$ & L$\mathrm{^{b}}$ & P$\mathrm{^{c}}$ & C$\mathrm{^{c}}$ \\
\hline
\multirow{2}*{0.0} &  91          &  4  & 2.1 & - & 6 & 0.6 & - \\
                     &  5           &  9  & -   & 2.0 & 31 & - & 7.6 \\
\hline
\multirow{2}*{$-1.5$} & 10.2        & 9   & 4.8 & - & 31 & 3.2 & - \\
                      & 29.8        & 6   & -  & 1.3 & 20 & -  & 4.9 \\
\hline
\multicolumn{8}{l}{$\mathrm{^{a}}$Detection threshold as listed in Table~\ref{tab:coverage} for P- and C-band,} \\
\multicolumn{8}{l}{\hspace{0.3cm}scaled to L-band using the listed spectral index.} \\
\multicolumn{8}{l}{$\mathrm{^{b}}$Number of detected bursts at L-band.} \\
\multicolumn{8}{l}{$\mathrm{^{c}}$Number of expected bursts given the L-band detections, the}\\
\multicolumn{8}{l}{\hspace{0.3cm}threshold, and the number of hours observed in each band.} \\
\end{tabular}
\end{table*}

\begin{table}
\centering
\caption{\label{tab:bright-ten-bursts}Dispersion measures as found by running {\tt DM-phase} \citep{seymour_2019_ascl} on ten bright, multi-component bursts detected throughout the campaign.}
\begin{tabular}{cc}
\hline
Burst ID & measured DM       \\ \hline
B06-o8   & $411.886\pm0.277$ \\
B08-o8   & $410.935\pm0.253$ \\
B12-o8   & $411.189\pm0.350$ \\
B13-o8   & $410.774\pm0.286$ \\
B18-wb   & $411.334\pm0.341$ \\
B19-wb   & $410.833\pm0.335$ \\
B20-wb   & $410.621\pm0.580$ \\
B29-wb   & $410.735\pm0.558$ \\
B34-wb   & $411.053\pm0.265$ \\
B42-wb   & $410.408\pm0.426$ \\ \hline
\end{tabular}
\end{table}

\bibliography{bibtex}{}
\end{document}